\documentclass[%
 aps,
 jmp,%
 amsmath,amssymb,
 reprint,%
]{revtex4-1}

\usepackage{graphicx}
\graphicspath{{figures_png/}}
\usepackage{dcolumn}
\usepackage{bm}
\usepackage{subfigure}

\usepackage{soul}
\usepackage{amsmath,amssymb,amsfonts}
\usepackage{graphics}
\usepackage{color}
\usepackage{xcolor}
\definecolor{rev1}{rgb}{0,0,0}

\usepackage{algorithm}
\usepackage{algorithmic}
\usepackage{enumitem}
\setlist[itemize]{leftmargin=*}
\setlist[enumerate]{leftmargin=*}

\usepackage{comment}
\usepackage{hyperref}
\usepackage{lipsum}
\usepackage{tabularx}
\usepackage{multirow}
\usepackage{mathrsfs}
\usepackage{stackengine}
\newcommand{\overbar}[1]{\mkern 1.5mu\overline{\mkern-1.5mu#1\mkern-1.5mu}\mkern 1.5mu}

\usepackage{accents}
\newcommand{\ubar}[1]{\underaccent{\bar}{#1}}

\begin{document}



\title{Frame invariant neural network closures for Kraichnan turbulence}


\author{Suraj Pawar}
\affiliation{ 
School of Mechanical \& Aerospace Engineering, Oklahoma State University, Stillwater, OK 74078, USA.
}%

\author{Omer San}%
 \email{osan@okstate.edu}
\affiliation{ 
School of Mechanical \& Aerospace Engineering, Oklahoma State University, Stillwater, OK 74078, USA.
}%

\author{Adil Rasheed}%
\affiliation{ 
	Department of Engineering Cybernetics, Norwegian University of Science and Technology, N-7465, Trondheim, Norway.
}%

\author{Prakash Vedula}%
\affiliation{ 
School of Aerospace \& Mechanical Engineering, The University of Oklahoma, Norman, OK 73019, USA.
}%



\date{\today}

\begin{abstract}
Numerical simulations of geophysical and atmospheric flows have to rely on parameterizations of subgrid scale processes due to their limited spatial resolution. Despite substantial progress in developing parameterization (or closure) models for subgrid scale (SGS) processes using physical insights and mathematical approximations, they remain imperfect and can lead to inaccurate predictions. In recent years, machine learning has been successful in extracting complex patterns from high-resolution spatio-temporal data, leading to improved parameterization models, and ultimately better coarse grid prediction. However, the inability to satisfy known physics and poor generalization hinders the application of these models for real-world problems. In this work, we propose a frame invariant closure approach to improve the accuracy and generalizability of deep learning-based subgrid scale closure models by embedding physical symmetries directly into the structure of the neural network. Specifically, we utilized specialized layers within the convolutional neural network in such a way that desired constraints are theoretically guaranteed without the need for any regularization terms. We demonstrate our framework for a two-dimensional decaying turbulence test case mostly characterized by the forward enstrophy cascade. We show that our frame invariant SGS model (i) accurately predicts the subgrid scale source term, (ii) respects the physical symmetries such as translation, Galilean, and rotation invariance, and (iii) is numerically stable when implemented in coarse-grid simulation with generalization to different initial conditions and Reynolds number. This work builds a bridge between extensive physics-based theories and data-driven modeling paradigms, and thus represents a promising step towards the development of physically consistent data-driven turbulence closure models.
\end{abstract}


\maketitle


\section{Introduction}

Computational modeling of turbulent flows remains a key issue in many engineering and geophysical applications. Turbulence involves a wide range of spatio-temporal scales that makes the direct numerical simulation (DNS) computationally infeasible for many complex systems. Coarse-graining approaches like large eddy simulation (LES) alleviate the computational burden by resolving only large and intermediate scales of the flow. The non-linearity of the Navier-Stokes equations introduces a subgrid scale (SGS) closure problem in LES which can be addressed via modeling of unresolved scales on the resolved flow quantities. The choice of the SGS model directly affects the accuracy of LES-based solution, and, therefore, the SGS modeling has been an active area of research for the past few decades \cite{pope2000turbulent, sagaut2006large, berselli2006mathematics,meneveau2000scale}. The development of SGS models has largely been driven by physical insights, mathematical considerations, and often problem-specific intuition. More recently, the availability of data from observations and high-resolution simulation along with advances in hardware and algorithms has fuelled interest in the development of data-driven turbulence models \cite{duraisamy2019turbulence, beck2021perspective, brunton2019machine, brenner2019perspective}.  

The initial efforts towards data-driven SGS modeling include training a neural network to predict computationally expensive SGS model for channel flow \cite{sarghini2003neural} with the aim to speed-up LES. Similar frameworks includes applying neural network to determine the eddy-viscosity of the dynamic Smagorinsky model \cite{pal2019deep,pawar2020apriori}, SGS model classification and blending \cite{maulik2019sub}, and reinforcement learning to predict SGS dissipation coefficient \cite{novati2021automating}. Deep learning (DL) has been applied to discovering new SGS models from the DNS data without any assumption of prior structural or functional form of the model \cite{gamahara2017searching,maulik2019subgrid,xie2019artificial,wang2021artificial,rasp2018deep,gagne2020machine,pawar2021nonintrusive}. The data-driven approach that employs convolutional neural network for learning the SGS model has also been used for different problems like two-dimensional decaying turbulence \cite{pawar2020apriori, pawar2021data, guan2021stable}, three-dimensional decaying homogeneous isotropic turbulence \cite{beck2019deep}, momentum forcing in ocean models \cite{bolton2019applications}, and subgrid-scale scalar flux modeling \cite{frezat2021physical}. Moreover, neural networks have also been utilized to learn the optimal map between filtered and unfiltered variables in the approximate deconvolution framework for SGS modeling \cite{maulik2018data,yuan2020deconvolutional}. Apart from SGS closure modeling, machine learning (ML) and in particular DL is being increasingly applied for different problems in fluid mechanics, like superresolution of turbulent flows \cite{fukami2019super, kim2021unsupervised}, Reynolds-Average Navier-Stokes (RANS) closure modeling \cite{wu2018physics,parish2016paradigm,srivastava2021generalizable}, and reduced-order modeling \cite{xiao2019domain,murata2020nonlinear,kaptanoglu2021physics,ahmed2021closures}.  
 
Despite their early success, ML models are faced with an array of challenges, such as poor generalization, lack of interpretability, and in some cases, violation of the known governing laws of the physical systems. For example, the SGS model derived through supervised learning may be numerically unstable, and diverge from the original trajectory, and this issue is exposed in many studies \cite{maulik2019subgrid, wu2019reynolds, beck2019deep, nadiga2007instability}. These issues can be addressed by leveraging our prior knowledge about the physical systems into an ML model. Readers are referred to recent review articles on physics-informed machine learning \cite{karniadakis2021physics,kashinath2021physics} that detail different methods of incorporating physics into ML models and discuss the capabilities and limitations of these methods. In the context of SGS modeling, there are many ways to embed physical constraints into the ML model. One such method for constructing a robust and generalizable SGS model is through the selection of suitable non-dimensionalized input and output quantities of the ML model to ensure that the known symmetries are respected \cite{prakash2021invariant}. 
Another class of methods pertains to the customized neural network architectures that encode the prior physical or mathematical knowledge as hard constraints. Some of the examples of this methods applied in fluid dynamics are tensor basis neural network \cite{ling2016reynolds}, transformation invariant neural network \cite{frezat2021physical}, physics-embedded neural network \cite{mohan2020embedding}, spatial transformer \cite{chattopadhyay2021towards,kneer2021symmetry}, and equivariant networks \cite{wang2021incorporating, siddani2021rotational}.       

In this work, we address the challenges associated with data-driven turbulence modeling by introducing a frame invariant convolutional neural network (FI-CNN) for SGS closure model discovery. Specifically, we select model inputs that are Galilean invariant, and replace the convolution operation with group convolutions \cite{cohen2016group, weiler20183d} to embed rotation invariance. Therefore, the FI-CNN preserves various symmetries, including translation, Galilean, and rotation both during training and inference. This makes the FI-CNN framework physically consistent and robust to extrapolation, and consequently, it produces accurate and stable results in their \textit{a posteriori} deployment. We demonstrate our framework for two-dimensional turbulence which is often used as a prototypical test case for large-scale geophysical flows \cite{boffetta2012two,bouchet2012statistical}. Although we focus on SGS closure model development in this study, this framework has a promising application for many scientific problems where physical symmetries are very common. For example, there are several invariant finite-difference schemes based on equivariant moving frames that preserve Lie symmetries that have been developed for the solution of partial differential equations (PDEs) via consideration of modified forms of the underlying PDEs \cite{ozbenli2020construction,ozbenli2017high,ozbenli2017numerical}. These symmetries can be exploited along with data-driven discretization \cite{bar2019learning} to design numerical schemes that are more accurate than their non-invariant counterpart. 

This paper is organized as follows. In Sec~\ref{sec:sgs}, the symmetries of Navier-Stokes equations and the SGS closure modeling problem for two-dimensional turbulence is introduced. The detailed procedure on how to embed frame symmetries, including translation, Galilean, and rotation invariance within the CNN is provided in Sec~\ref{sec:fi_cnn}. In Sec~\ref{sec:data}, the details on data generation and training are discussed. The performance of the FI-CNN in the \textit{a priori} and \textit{a posteriori} settings along with a detailed discussion of the results are presented in Sec~\ref{sec:results}. Finally, the concluding remarks and summary of the work are given in Sec~\ref{sec:conclusion}.

\section{SGS Closure Modeling} \label{sec:sgs}

\subsection{Symmetries of Navier-Stokes Equations}
The Navier-Stokes equations governing incompressible fluid flows can be written in primitive variable (velocity–pressure) form as 
\begin{align} 
    \nabla \cdot \boldsymbol{u} &= 0, \\
    \frac{\partial \boldsymbol{u}}{\partial t} + \boldsymbol{u} \cdot \nabla \boldsymbol{u} &= -\frac{1}{\rho} \nabla p + \nu \nabla^2 \boldsymbol{u},\label{eq:ns}
\end{align}
where $\boldsymbol{u}$ is the velocity, $p$ is the pressure, $\rho$ is the density, and $\nu$ is the kinematic viscosity of the fluid. The governing equations for LES (also called as the filtered Navier-Stokes equations) are obtained by applying a spatial filter operation and it can be written as follows

\begin{align} 
    \nabla \cdot \bar{\boldsymbol{u}} &= 0, \\
    \frac{\partial \bar{\boldsymbol{u}}}{\partial t} + \bar{\boldsymbol{u}} \cdot \nabla \bar{\boldsymbol{u}} &= -\frac{1}{\rho} \nabla \bar{p} + \nu \nabla^2 \bar{\boldsymbol{u}} + \nabla \cdot \underbrace{(\bar{\boldsymbol{u}} \bar{\boldsymbol{u}} - \overbar{\boldsymbol{u} \boldsymbol{u}})}_{\boldsymbol{\tau}(\boldsymbol{u}, \boldsymbol{u})} ,\label{eq:fns}
\end{align}
where the overbar is used to denote the filtered variables and $\boldsymbol{\tau}(\boldsymbol{u}, \boldsymbol{u})$ is subgrid-scale stress tensor. The problem of determining subgrid-scale stress tensor $\boldsymbol{\tau}$ using the filtered variables is called the subgrid scale closure problem in LES. 

There are many possible SGS closure models and any mathematical, physical constraints will lead to a specific type of SGS model. Requirements such as frame invariance, realizability, finite kinetic energy can act as guiding principles for a satisfactory SGS closure model, and readers are referred to \cite{berselli2006mathematics} for more details. The frame invariance constraint on the SGS model is derived by enforcing the symmetry of the original Navier-Stokes equations \cite{sagaut2006large, pope2000turbulent, oberlack1997invariant,berselli2006mathematics} upon the filtered Navier-Stokes equations with the SGS closure model. Let $G$ denote a group of transformation acting on space-time functions $\boldsymbol{u}(\boldsymbol{x},t)$. We say that the group $G$ is a symmetry group of the Navier-Stokes equations if, for all $\boldsymbol{u}$ which are solutions of the Navier-Stokes equations, and all $g \in G$, the function $g \boldsymbol{u}$ is also a solution \cite{frisch1995turbulence}.
The frame invariance constraint involves preservation of the symmetry property of the original Navier-Stokes equations to translation, Galilean, and rotation transformations, and they can be written as follows 
\begin{itemize}
    \item Space-translation: $g_{\boldsymbol{\delta}}^{\text{space}} \boldsymbol{u}(\boldsymbol{x},t) = \boldsymbol{u}(\boldsymbol{x} - \boldsymbol{\delta},t),~ \forall \boldsymbol{\delta} \in \mathbb{R}^3 $, where $g_{\boldsymbol{\delta}}^{\text{space}}$ is the space-translation operator with the arbitrary displacement $\boldsymbol{\delta}$.
    \item Galilean transformation: $g_{\boldsymbol{\alpha}}^{\text{Gal}} \boldsymbol{u}(\boldsymbol{x},t) = \boldsymbol{u}(\boldsymbol{x}-\boldsymbol{\alpha}t,t) + \boldsymbol{\alpha} ,~ \forall {\boldsymbol{\alpha}} \in \mathbb{R}^3 $, where $g_{\boldsymbol{\alpha}}^{\text{Gal}}$ is the Galilean operator and $\boldsymbol{\alpha}$ is a fixed but arbitrary constant vector. 
    \item Space-rotations: $g_{\boldsymbol{A}}^{\text{rot}} \boldsymbol{u}(\boldsymbol{x},t) = \boldsymbol{A}\boldsymbol{u}(\boldsymbol{A}^{-1}\boldsymbol{x},t)$, where $g_{\boldsymbol{A}}^{\text{rot}}$ is the rotation operation and $\boldsymbol{A} \in SO(3)$. 
\end{itemize}
Imposing the symmetry preservation constraint give some structure to the the SGS model, and this insights have been extensively used in turbulence models \cite{speziale1985galilean, sagaut2006large, berselli2006mathematics}. We exploit these symmetries as physical constraints while building a frame invariant data-driven SGS model.

\subsection{Two-dimensional Turbulence}
In this work, we are interested in the SGS modeling for two-dimensional turbulence that is usually applied for modeling geophysical flows in the atmosphere and ocean \cite{bouchet2012statistical,boffetta2012two} where rotation and stratification dominate, and the most efficient way to model it is using the vorticity transport equation. Taking the curl of Eq.~\ref{eq:ns} yields the Navier-Stokes equations in vorticity-velocity formulation, and, for incompressible fluid flows, it can be written as follows
\begin{align} 
    \frac{\partial \omega}{\partial t} + (\boldsymbol{u} \cdot \nabla ) \omega &=  \nu \nabla^2 \omega,\label{eq:ns_ws}
\end{align}
where $\omega$ is the vorticity, and for two-dimensional flows, we have $\omega={\partial v}/{\partial x} - {\partial u}/{\partial y}$. A scalar function called the streamfunction is defined in such a way that the continuity equation is satisfied if the velocity expressed in terms of the streamfunction is substituted in the continuity equation. This leads to the definition of velocity in terms of the streamfunction as follows
\begin{equation}
    u = \frac{\partial \psi}{\partial y}, \quad v = -\frac{\partial \psi}{\partial x},
\end{equation}
where $\psi$ is the streamfunction. The Poisson equation relating the vorticity and streamfunction is obtained by substituting the above velocity components in the definition of vorticity. Thus, we have 
\begin{equation}
    \nabla^2 \psi = -\omega.
\end{equation}
It is convenient to write Eq.~\ref{eq:ns_ws} in the vorticity-streamfunction formulation as follows
\begin{align}
     {\frac{\partial \omega}{\partial t}+J(\omega,\psi) = \frac{1}{\text{Re}}\nabla^2\omega}, \label{eq:ws}\\
    {J(\omega, \psi) = \frac{\partial \omega}{\partial x} \frac{\partial \psi}{\partial y} - \frac{\partial \omega}{\partial y} \frac{\partial \psi}{\partial x},}
\end{align}
where $J(\cdot , \cdot)$ is the Jacobian (or the nonlinear term), and Re is the Reynolds number of the flow. The above equation is also called the vorticity transport equation.


The filtered Navier-Stokes equations for two-dimensional turbulence is obtained by applying a spatial filtering operation to Eq~\ref{eq:ws} as follows
\begin{equation}
    \frac{\partial \overbar{\omega}}{\partial t} + \overbar{J({\omega},{\psi})} = \frac{1}{\text{Re}}\nabla^2\overbar{\omega}.
\end{equation}
The above equation can be rewritten as 
\begin{equation}
    \frac{\partial \overbar{\omega}}{\partial t} + {J(\overbar{\omega},\overbar{\psi})} = \frac{1}{\text{Re}}\nabla^2\overbar{\omega} + \Pi, \label{eq:les_w}
\end{equation}
where the overbar quantities represent filtered variables and are evolved on a grid that is significantly coarse compared to the DNS resolution. The effect of the unresolved scales due to truncation of high wavenumber flow scales is encapsulated in a subgrid scale (SGS) source term $\Pi$ and must be modeled solely based on the resolved variables $(\overbar{\omega}, \overbar{\psi})$. Mathematically, the true SGS source term $\Pi$ can be expressed as 
\begin{equation}
    {\Pi} = J(\overbar{\omega},\overbar{\psi}) - \overbar{J(\omega, \psi)}.
\end{equation}

The functional and structural models are the most commonly used approaches for modeling the SGS closure term in LES of turbulent flows \cite{sagaut2006large}. The functional models are based on the concept of eddy viscosity where the effect of unresolved scales are approximated by artificial dissipation \cite{smagorinsky1963general,leith1971atmospheric}. The functional models can be further improved by dynamic adaptations of the coefficients that control the dissipation of the model and are determined adaptively by the use of a low-pass spatial test filter \cite{germano1991dynamic, lilly1992proposed, frederiksen2006dynamical}. Although the dynamic formulation allows for spatial and temporal variation of coefficients in the eddy viscosity model, the ensemble averaging procedure does not allow for true back-scattering in order to limit the growth of numerical instabilities during the \textit{a posteriori} testing \cite{kirkil2012implementation,iliescu2004backscatter}. The structural models on the other hand aim at obtaining an accurate approximation of the SGS term and are based on the approximate deconvolution procedure \cite{stolz2001approximate,san2013approximate} and scale-similarity arguments \cite{bardina1980improved}. Scale-similarity models address the SGS closure term by extrapolation from the smallest resolved scales to unresolved scales and have found to be the most accurate in \textit{a priori} testing \cite{sagaut2006large, sarghini1999scale}. However, numerical instabilities have been reported with scale-similarity models, and this has led to development of many mixed models with additional eddy viscosity term for stability reasons \cite{layton2003simple, liu1994properties, maulik2017stable}. More recently, data-driven methods are emerging as a new paradigm to build turbulence closure models by extracting information from the data, and are seen as the potential applications to address the limitations of existing SGS models \cite{duraisamy2019turbulence, beck2021perspective, duraisamy2021perspectives}.


\section{Frame invariant SGS closure model} \label{sec:fi_cnn}
In this work, we consider the frame invariance property that must be satisfied by any SGS model and demonstrate how to include them within a neural network as hard constraints. The SGS source term $\Pi$ is approximated using a neural network as shown below
\begin{equation}
    \widetilde{\Pi} \approx \mathcal{M}(\bar{\omega}, \bar{\psi}),
\end{equation}
where $\mathcal{M}$ is a neural network-based model, and $\widetilde{\Pi}$ is the approximation of true SGS source term $\Pi$. We remark here that the vorticity is defined using the spatial derivative of the velocity field, and, therefore it is invariant to Galilean transformations. Additionally, the streamfunction is computed using the vorticity, and therefore both the inputs to our model are Galilean invariant. We now discuss how to embed the translation and rotational invariance/symmetry properties into the neural network-based model.  


\subsection{Translation invariance}\label{sec:traslation}
In this work, we employ the convolutional neural network (CNN) for learning the SGS closure model based on filtered vorticity and streamfunction as the model inputs. The CNN is an attractive choice for high-dimensional data and it does not suffer from the curse of dimensionality due to its weight-sharing feature. The CNN is composed of many convolutional layers and each of the layers is parameterized by filters, also called kernels, that has to be learned through training. Let $\ubar{f},\ubar{k} : \mathbb{R}^2 \rightarrow \mathbb{R}^{N_c}$ be vector-valued two-dimensional features and kernel, i.e., $\ubar{f}=(f_1,\cdots,f_{N_c})$ and $\ubar{k}=(k_1,\cdots,k_{N_c})$, then the convolutional operation can be defined as 
\begin{equation}
    (\ubar{k} \star \ubar{f})(\mathbf{x}) = \sum_{c=1}^{N_c} \int_{\mathbb{R}^2}k_c(\mathbf{x} - \mathbf{x^\prime})f_c(\mathbf{x^\prime})d\mathbf{x^\prime},
\end{equation}
where $\mathbf{x^\prime}$ is a dummy variable spanning over $\mathbb{R}^2$ space. The convolutional layer maps a feature vector $\ubar{f}^{(l-1)}:\mathbb{R}^2 \rightarrow \mathbb{R}^{N_{l-1}}$ with $N_{l-1}$ channels to feature vector $\ubar{f}^{(l)}:\mathbb{R}^2 \rightarrow \mathbb{R}^{N_{l}}$ using a set of $N_l$ kernels $\mathbf{k}^{(l)}:=(\ubar{k}_1^{(l)}, \cdots, \ubar{k}_{N_l}^{(l)})$ and this operation can be defined as 
\begin{equation}
    \ubar{f}^{(l)} = \zeta(\mathbf{k}^{(l)} \star \ubar{f}^{(l-1)}) := \zeta(\ubar{k}_1^{(l)} \star \ubar{f}^{(l-1)}, \cdots , \ubar{k}_{N_l}^{(l)} \star \ubar{f}^{(l-1)}),
\end{equation}
where $\zeta$ is an activation function. The parameters of the kernel are shared for the whole image as the kernel is convolved relative to the position about $\mathbf{x}$ and this aspect of the relative motion makes the CNN translation invariant. Although we present the convolution operation with continuous kernels, convolutional layers are equipped with discretized-filtering operations in their practical implementation. From here on, we refer to the model build using convolutional layers and nonlinear activation function as $\mathcal{M}_{\text{CNN}}$. The inputs to our model are the vorticity and streamfunction and the output is the SGS source term. Therefore, the learning map for $\mathcal{M}_{\text{CNN}}$ can be expressed as 
\begin{equation} \label{eq:map_cnn}
    \mathcal{M}_{\text{CNN}}:\{\bar{\omega}, \bar{\psi}\} \in \mathbb{R}^2 \rightarrow \mathbb{R}^2 \mapsto \{\widetilde{\Pi}\} \in \mathbb{R}^2 \rightarrow \mathbb{R}^1,
\end{equation}
where $\widetilde{\Pi}$ is the predicted SGS source term.

\subsection{Rotation invariance}\label{sec:rotation}
The rotational invariance of the SGS model requires that it maps as a tensor under the coordinate rotation \cite{sagaut2006large}. As discussed in Section~\ref{sec:traslation}, the CNN is often invariant to only translation and not for other groups of transformations. However, there are recent developments on this front to exploit polar mapping of input images to convert rotation to translation \cite{kim2020cycnn}.  
In this work, we apply the group equivariant convolutions within the \emph{E(2)-CNN} framework \cite{weiler2019general} for embedding rotational symmetry. The first roto-translation equivariant CNN was called the group convolutional neural network (GCNN) and it considered the rotations by multiples of $\pi/2$ \cite{cohen2016group}. The GCNN was further augmented by defining filters in terms of the steerable basis that are equivariant to rotations by multiples of $2\pi/N$, with $N>4$ \cite{weiler2018learning}. The \emph{E(2)-CNN} library is based on the framework of steerable CNNs \cite{cohen2016steerable,weiler20183d} and it has different options for the group that takes the form of the semi-direct group $H=\mathbb{R}^2 \rtimes G$ where the group $G \leq O(2)$ (here $O(2)$ is the group of continuous rotations and reflections). For example, the group $H =\mathbb{R}^2 \rtimes SO(2)=SE(2)$ is the semi-direct product of the group of planar translations  $\mathbb{R}^2$ and continuous rotations $SO(2)$. In this work, we utilize the cyclic group $G=C_N$ containing the discrete rotations of $2\pi/N$ (i.e., $H=\mathbb{R}^2 \rtimes C_N$). For a large value of $N$, the difference between continuous rotations and discrete rotations is indistinguishable due to space discretization.  

A full understanding of the steerable CNNs requires some knowledge of the group representation theory, but the implementation of the steerable CNNs is similar to ordinary CNNs. Readers are suggested to read \citeauthor{weiler2019general} \cite{weiler2019general} and references therein for a more comprehensive discussion on the general framework of steerable CNNs. Here, we briefly explain the $G$-equivariant convolutions. A $G$-convolution between a vector-valued two-dimensional image $\ubar{f}:\mathbb{R}^2\rightarrow \mathbb{R}^{N_c}$ and a filter $\ubar{k}:\mathbb{R}^2\rightarrow \mathbb{R}^{N_c}$ where $\ubar{f}=(f_1,\cdots,f_{N_c})$ and $\ubar{k}=(k_1,\cdots,k_{N_c})$ can be expressed as follows
\begin{equation}
    (\ubar{k} \tilde{\star} \ubar{f})(g) = \sum_{c=1}^{N_c} \int_{\mathbb{R}^2}k_c(g^{-1}\mathbf{x^\prime})f_c(\mathbf{x^\prime})d\mathbf{x^\prime},
\end{equation}
where $g=(\mathbf{x},\theta)\in H=\mathbb{R}^2 \rtimes C_N$, $\mathbf{x^\prime} \in \mathbb{R}^2$, and $\tilde{\star}$ denotes the group correlation operation under joint translation and rotation. This operation corresponds to lifting of the data on two-dimensional space to the data that lives on a three-dimensional position orientation space $H$. The first layer maps a two-dimensional image $\ubar{f}^{(l-1)}:\mathbb{R}^2\rightarrow \mathbb{R}^{N_{l-1}}$ with $N_{l-1}$ channels at $(l-1)$th layer to $H$ vector image $\ubar{F}^{(l)}:H \rightarrow \mathbb{R}^{N_{l}}$ using a set of $N_l$ kernels $\mathbf{k}^{(l)}:=(\ubar{k}_1^{(l)}, \cdots, \ubar{k}_{N_l}^{(l)})$ as follows
\begin{equation}
    \ubar{F}^{(l)} = \zeta(\mathbf{k}^{(l)} \tilde{\star} \ubar{f}^{(l-1)}) := \zeta(\ubar{k}_1^{(l)} \tilde{\star} \ubar{f}^{(l-1)}, \cdots , \ubar{k}_{N_l}^{(l)} \tilde{\star} \ubar{f}^{(l-1)}).
\end{equation}
Since the $\ubar{F}$ is a function on $H$, the filters from the second layer onward should also be functions on $H$. The subsequent group convolutions are defined as \cite{weiler2018learning,bekkers2018roto} 
\begin{equation}
    (\ubar{K} \tilde{\star} \ubar{F})(g) = \sum_{c=1}^{N_c} \int_{H}K_c(g^{-1}h)F_c(h)dh.
\end{equation}
A group convolution layer is defined by a set of $H$ kernels $\mathbf{K}:=(\ubar{K}_1^{(l)},\cdots\ubar{K}_{N_l}^{(l)})$ that maps $\ubar{F}^{(l-1)}$ with $N_{l-1}$ channels to $\ubar{F}^{(l)}$ with $N_{l}$ channels as shown below
\begin{multline}
    \ubar{F}^{(l)} = \zeta(\mathbf{K}^{(l)} \tilde{\star} \ubar{F}^{(l-1)}) := \zeta(\ubar{K}_1^{(l)} \tilde{\star} \ubar{F}^{(l-1)}, \cdots , \\
    \ubar{K}_{N_l}^{(l)} \tilde{\star} \ubar{F}^{(l-1)}).
\end{multline}
Finally, the feature field at the last layer can be synthesized from $H$ space to $\mathbb{R}^2$ space. The user interface of the \emph{E(2)-CNN} library \cite{weiler2019general} hides most of the intricacies of group theory, solutions of the steerable kernels space constraints, and requires users to specify only the transformation laws of the feature spaces. We use the regular representation for all hidden layers and the action of regular representation is given by permutation matrices (Appendix B in \cite{weiler2019general}). From here on, the model built using the equivariant CNN is called as $\mathcal{M}_{\text{FI-CNN}}$. The learning map for $\mathcal{M}_{\text{FI-CNN}}$ is same as the $\mathcal{M}_{\text{CNN}}$ given in Eq.~\ref{eq:map_cnn}.

\section{Data Generation and Training} \label{sec:data}
The parameters of the neural network based SGS models are learned through supervised training that requires a set of labeled inputs and outputs, usually obtained from direct numerical simulation (DNS). The dataset should encompass a range of dynamics that is expected to be reproduced by the SGS model. The data for training is generated from DNS of two-dimensional Kraichnan turbulence in a doubly periodic square domain with $L_x \times L_y = [0,2\pi] \times [0,2\pi]$, and the domain is discretized using $2048^2$ degrees of freedom. Our DNS solver  is based on a second-order accurate energy-conserving Arakawa scheme \cite{arakawa1997computational} for the nonlinear Jacobian and second-order accurate finite-difference scheme for the Laplacian of the vorticity. The elliptic equation for the relationship between the streamfunction and vorticity is solved using a second-order accurate FFT-based Poisson solver, and the time integration is performed with a third-order accurate Runge-Kutta method. The vorticity distribution at the start of the simulation is initialized based on the energy spectrum given by \cite{orlandi2000fluid}
\begin{equation}
    E(k) = Ak^4 \text{exp}\bigg(-\bigg(\frac{k}{k_p}\bigg)^2 \bigg),
    \label{eq:en0}
\end{equation}
where $A=4k_p^{-5}/3\pi$ and $k=|\mathbf{k}|=\sqrt{k_x^2+k_y^2}$. For our numerical experiments, we use $k_p=10$. The initial vorticity distribution in Fourier space is obtained through the introduction of random phase as follows
\begin{equation}
    \tilde{\omega}(\mathbf{k}) = \sqrt{\frac{k}{\pi}E(k)}~\text{e}^{\mathbf{i}\xi(\mathbf{k})},
\end{equation}
where the phase function is given by $\xi(\mathbf{k})=\phi(\mathbf{k}) + \eta(\mathbf{k})$. Here, $\phi(\mathbf{k})$ and $\eta(\mathbf{k})$ are independent random values chosen in $[0,2\pi]$ at each grid point in the first quadrant of the $k_x-k_y$ plane (i.e., $k_x,k_y \ge 0$). The phase function for other quadrants is obtained through conjugate relations as follows
\begin{align}
    \xi(\mathbf{k}) &= -\phi(\mathbf{k}) + \eta(\mathbf{k}) ~\text{for}~ k_x < 0 ~\text{and}~ k_y \ge 0, \\
    \xi(\mathbf{k}) &= -\phi(\mathbf{k}) - \eta(\mathbf{k}) ~\text{for}~ k_x < 0 ~\text{and}~ k_y < 0, \\
    \xi(\mathbf{k}) &= \phi(\mathbf{k}) - \eta(\mathbf{k}) ~\text{for}~ k_x \ge 0 ~\text{and}~ k_y < 0, \\
\end{align}
Further details on the problem setup and the numerical schemes can be found in our previous work \cite{maulik2017stable}. Different realizations of the initial vorticity field can be obtained by using different phase functions with a different seed for random value generation.

\begin{figure*}[htbp]
\centering
\mbox{\subfigure{\includegraphics[width=0.8\textwidth]{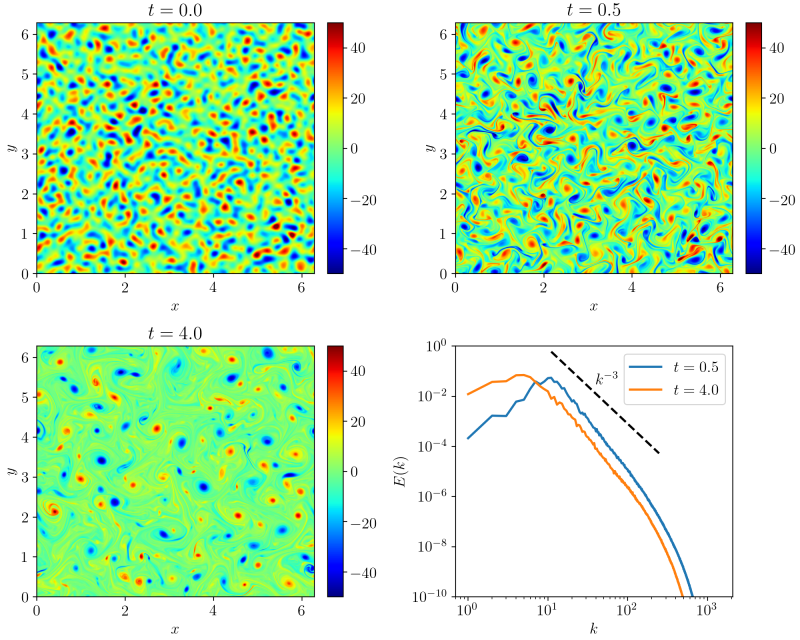}}}
\caption{Visualization of the vorticity field and energy spectrum at different time instances for Re = 16000 with grid resolution $2048 \times 2048$.}
\label{fig:flow}
\end{figure*}

The DNS is performed from time $t=0$ to $t=4$ with the time step $\Delta t = 5 \times 10^{-4}$. In the Kraichnan turbulence problem, the initial vorticity field is dominated by a population of vortices and small-scale structure starts appearing as the flow evolves. The initial spin-up time from $t=0$ to $0.5$ is neglected and we start collecting the data for training from time $t=0.5$. From time $t\approx0.5$, the flow has started following Kraichnan–Batchelor–Leith (KBL) theory \cite{kraichnan1967inertial,batchelor1969computation,leith1971atmospheric} of energy cascade where energy is transferred from the smaller scales to the larger scales. From time $t\approx0.5$ onward, large coherent vortices start emerging through vortex merging mechanism and viscous dissipation of small-scale structures. The vorticity field and angle-averaged energy spectrum are displayed in Fig.~\ref{fig:flow} and we can see that the energy spectrum has started exhibiting $k^{-3}$ scaling from approximately $t=0.5$.

The data for training a neural network-based SGS model is stored at every $20 \Delta t$, i.e., we have 350 snapshots of the vorticity and streamfunction between $t=0.5$ to $4.0$. We emphasize here that the neural network-based SGS model is trained only for a single Reynolds number Re = 16000 and we assess the performance of the model for Reynolds number up to Re = 128000. The filtered DNS data for training is obtained by first applying a Gaussian filter transfer function to the DNS data and then coarse-graining the filtered solution to the LES grid \cite{zanna2020data,guan2021stable}. The Gaussian filter provides a smooth transition between resolved and subgrid scales and is also positive definite in physical and wave space \cite{piomelli1991subgrid, rogallo1984numerical}. Additionally, our numerical solver is in physical space, and therefore we select the Gaussian filter instead of a spectral cut-off filter. The coarse-grid level for LES is $256^2$ which corresponds to 64 times fewer spatial degrees of freedom compared to DNS.

\begin{figure}[htbp]
\centering
\mbox{\subfigure{\includegraphics[width=0.45\textwidth]{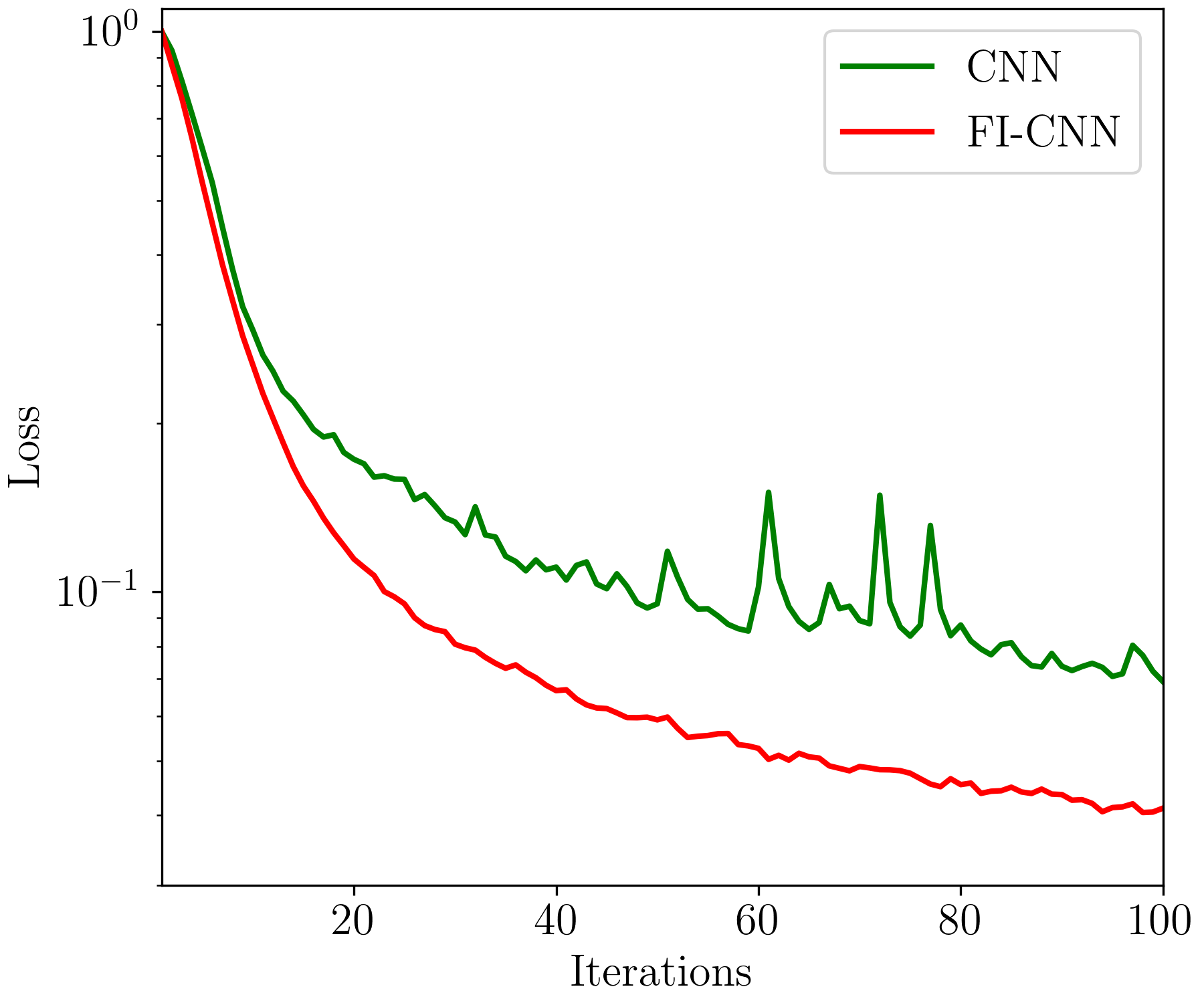}}}
\caption{History of the training loss versus iterations for $\mathcal{M}_{\text{CNN}}$ and $\mathcal{M}_{\text{FI-CNN}}$.}
\label{fig:loss}
\end{figure}

We do not pre-process the filtered DNS data before training as the DNS data is generated from a non-dimensionalized vorticity transport equation. The total data is divided into 80\% of the data for training and 20\% for the validation set. While the input and output of both $\mathcal{M}_{\text{CNN}}$ and $\mathcal{M}_{\text{FI-CNN}}$ are the same, the user needs to specify the type of representation for intermediate feature field while constructing an FI-CNN \cite{weiler2019general}, similar to the number of channels for CNN. We use the kernel size of $5 \times 5$, six hidden layers and ReLU activation function for the $\mathcal{M}_{\text{CNN}}$ and $\mathcal{M}_{\text{FI-CNN}}$. The number of channels for the CNN and FI-CNN models is set to 30 and 16, respectively. With these hyperparameters, the number of trainable parameters is roughly the same around $O(1.1 \times 10^5)$ for both models. Both the models are trained for 100 iterations using an Adam optimizer. Fig.~\ref{fig:loss} shows the history of training loss versus iterations for both neural network-based SGS models and we can observe that the loss for $\mathcal{M}_{\text{FI-CNN}}$ is almost one order magnitude less than the loss for $\mathcal{M}_{\text{CNN}}$. This can be attributed to rotational invariances incorporated in the $\mathcal{M}_{\text{FI-CNN}}$ against $\mathcal{M}_{\text{CNN}}$, which is only invariant to translation and Galilean transformation. For both neural network-based SGS models, we use the parameters (i.e., weights) corresponding to minimum validation loss obtained while training the neural network. 

\section{Numerical Results} \label{sec:results}
We first outline the numerical results of our framework in the \textit{a priori} settings where the neural network-based models are utilized in predicting the SGS source term. We analyze the capability of $\mathcal{M}_{\text{CNN}}$ and $\mathcal{M}_{\text{FI-CNN}}$ in incorporating the frame-invariance property over the testing data. Then, we present the results of \textit{a posteriori} LES coupled with neural network-based SGS models and evaluate their performance using numerous statistical metrics.  

\subsection{\textit{A Priori} investigation}
\begin{figure*}[htbp]
\centering
\mbox{\subfigure{\includegraphics[width=0.8\textwidth]{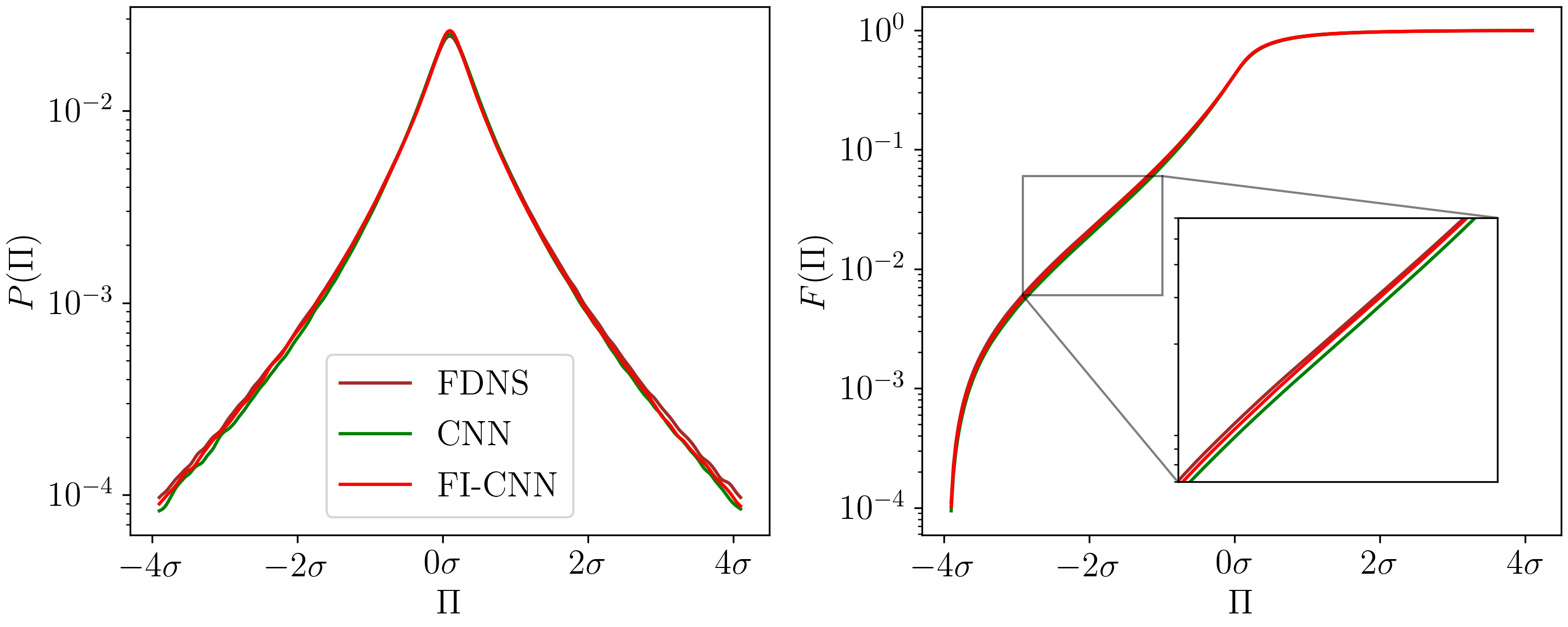}}}
\caption{Probability distribution function (left) and cumulative distribution function (right) of the SGS source term over the entire testing dataset. The testing dataset corresponds to 70 snapshots selected randomly between time $t=0.5$ to $t=4.0$ for the initial condition different from the one used in training.}
\label{fig:a_priori}
\end{figure*}

Here, we assess the performance of neural network-based models in predicting the SGS source term compared to the true SGS source term for the out-of-training data. The out-of-training data is obtained for a different initial condition and corresponds to 70 snapshots stored randomly between time $t=0.5$ to $t=4.0$. We remark here that the initial energy spectrum for the testing data is also given by Eq.~\ref{eq:en0} and the difference is due to a different phase function. Fig.~\ref{fig:a_priori} displays the probability distribution function and cumulative distribution function for the test data. There is a very good agreement between the true SGS source term and the predicted SGS source term from both models. However, we notice that the $\mathcal{M}_{\text{FI-CNN}}$ is more accurate near the tails of the distribution (Fig~\ref{fig:a_priori}, left) compared to $\mathcal{M}_{\text{CNN}}$. This difference is also observable in the cumulative distribution function of true and predicted SGS source terms and is highlighted in the zoom-in portion (Fig~\ref{fig:a_priori}, right). Based on these results, we may conclude that both neural network-based SGS model has learned the relationship between filtered quantities and the SGS source term. Both models are able to produce viable physical results for the completely unseen data with similar physics. 

Next, we evaluate neural network-based models in respecting rotational symmetry on the test data. Specifically, we perturbed the test data based on the rotation transformation, and generate multiple test datasets. Then, we compute the root mean squared error (RMSE) for each dataset, and calculate the expected value and variance for all test datasets. The rotation matrix $\mathbf{A}$ is sampled uniformly between $[0^\circ,360^\circ]$ in the multiple of $90^\circ$. If the rotation symmetry constraint is satisfied strictly, then the RMSE will be the same for each test dataset leading to zero variance for the RMSE metric. The results in Table~\ref{tab:rmse} demonstrate the remarkable ability of $\mathcal{M}_{\text{FI-CNN}}$ to respect the rotation symmetry in contrast to $\mathcal{M}_{\text{CNN}}$ which violates this symmetry. Furthermore, the RMSE for $\mathcal{M}_{\text{FI-CNN}}$ is one order of magnitude lower than $\mathcal{M}_{\text{CNN}}$ and is consistent with the training loss (Fig.~\ref{fig:loss}). The performance of both models is substantially accurate in terms of the Pearson's cross-correlation coefficient, with $\mathcal{M}_{\text{FI-CNN}}$ slightly better than $\mathcal{M}_{\text{CNN}}$. We note here that it is relatively straightforward to embed Galilean invariance constraint within neural network-based SGS model through intelligent selection of model inputs, and translation invariance through simple CNN. However, incorporating rotational symmetry in a neural network-based SGS model is more complex and requires special consideration. Although a relatively simple method like data augmentation can be utilized to impose the rotation symmetry as a soft constraint, it does not satisfy rotation invariance strictly \cite{frezat2021physical}. The strict enforcement of rotation symmetry is challenging and requires the use of tailored neural network architecture, such as equivariant CNN. 

\begin{table*}[]
\caption{Evaluation of the rotational symmetry constraints provided by $\mathcal{M}_{\text{CNN}}$ and $\mathcal{M}_{\text{FI-CNN}}$. The expected value and variance of
the root mean squared error on the SGS source term predicted by both models is computed from many realizations (20 ensembles) on the testing data. The testing dataset corresponds to 70 snapshots selected randomly for the initial condition different from the one used for training. The rotational angle $\mathbf{A}$ is sampled uniformly between $[0^\circ,360^\circ]$ in the multiple of $90^\circ$ and is used in the rotational operator $g_{\mathbf{A}}^{\text{rot}}$. The Pearson's cross-correlation coefficient between the predicted SGS source term and the filtered DNS solution is computed as $\mathcal{P}(X,Y) = \text{cov}(X,Y)/\sigma_X \sigma_Y$.}
\begin{tabular}{p{0.3\textwidth}p{0.25\textwidth}p{0.25\textwidth}}
\hline
\hline\noalign{\smallskip}
\multirow{2}{*}{\begin{tabular}[c]{@{}c@{}}\end{tabular}} 
Metric & \multicolumn{1}{l}{$\mathcal{M}_{\text{CNN}}$} & \multicolumn{1}{l}{$\mathcal{M}_{\text{FI-CNN}}$}\\ \noalign{\smallskip}
\hline\noalign{\smallskip}
$\text{E}[\mathcal{L}]$ & 10.6941 & 7.3462 \\ \noalign{\smallskip}
$\sigma[\mathcal{L}]$  & 4.2442 $\times 10^{-2}$ & 5.6587 $\times 10^{-8}$ \\ \noalign{\smallskip}
$\mathcal{P}(X,Y)$ & 0.9600 & 0.9776 \\ \noalign{\smallskip}
\hline
\end{tabular}\label{tab:rmse}
\end{table*}

\subsection{\textit{A Posteriori} deployment}
We now evaluate the performance of neural network-based SGS models in the LES of Kraichnan turbulence. The spatial resolution for LES is reduced by a factor of eight in each direction and this gives us $256^2$ degrees of freedom. The time step for LES simulation is ten times larger compared to the DNS, i.e., $\Delta t_{\text{LES}}=5 \times 10^{-3}$. The performance of neural network-based SGS models is compared with the widely used dynamic Smagorinsky model (DSM) \cite{germano1991dynamic, lilly1992proposed}. The \textit{a posteriori} deployment is a rigorous task for any data-driven SGS model due to the presence of numerical instabilities, and the challenges and remedies have been highlighted in many studies \cite{maulik2019subgrid,maulik2019sub,maulik2020spatiotemporally,guan2021stable,beck2019deep,stoffer2021development,zhou2019subgrid}. For example, \citeauthor{maulik2019subgrid} \cite{maulik2019subgrid} and \citeauthor{zhou2019subgrid} \cite{zhou2019subgrid} achieved the stable LES results by truncating SGS source term corresponding to negative eddy viscosity. \citeauthor{stoffer2021development} \cite{stoffer2021development} attained stable \textit{a posteriori} results by resorting to artificially introducing additional dissipation (via eddy-viscosity models). \citeauthor{guan2021stable} \cite{guan2021stable} provided sufficient amount of data during training to obtain a stable \textit{a posteriori} results. While the exact reason for this behavior is unknown, several issues such as error accumulation, aliasing errors, numerical instability, extrapolation beyond the training data, chaotic nature of turbulence, presence of multiple attractors might be responsible for unstable \textit{a posteriori} simulation \cite{beck2019deep,stoffer2021development,beck2021perspective,nadiga2007instability}. 

From our \textit{a posteriori} simulation, it is revealed that $\mathcal{M}_{\text{CNN}}$ is unstable, while $\mathcal{M}_{\text{FI-CNN}}$ is able to produce a stable and physical solution without any kind of clipping or by adding artificial dissipation. We note here that, perhaps $\mathcal{M}_{\text{CNN}}$ can also achieve stable \textit{a posteriori} simulation, provided there is sufficient data available for training or some kind of post-processing is carried out for the predicted SGS source term. However, our main motivation in this work is to construct a physically consistent data-driven SGS model that can be trained in a data-sparse regime and is also stable in the \textit{a posteriori} simulation. We assess the performance of our \textit{a posteriori} simulation using several statistical metrics and compare it with the statistics from filtered DNS solution. The turbulent kinetic energy at time $t_k$ is computed as follows
\begin{equation}
    TKE = \lambda(u_f^2 + v_f^2(t_k)),
\end{equation}
where $u_f$ and $v_f$ are the fluctuating components of velocity given by 
\begin{align}
    u_f = \bar{u} - \lambda(\bar{u}), \\
    v_f = \bar{v} - \lambda(\bar{v}),
\end{align}
where $\lambda(a)$ represents the spatial average of the field variable $a$. The velocity $\bar{u}$, and $\bar{v}$ are computed by spectral differentiation of streamfunction as shown below
\begin{equation}
    \bar{u} = \frac{\partial \bar{\psi}}{\partial y}; \quad \bar{v} = -\frac{\partial \bar{\psi}}{\partial x}.
\end{equation}
The vorticity variance at each time step is computed as 
\begin{equation}
    \sigma^2 = \lambda((\bar{\omega} - \lambda(\bar{\omega}))^2).
\end{equation}

We compare the kinetic-energy spectra and the vorticity structure function at intermediate time $t=2.0$ and at final time $t=4.0$ with the $k^{-3}$ scaling which is observed in two-dimensional turbulence based on the classical KBL theory. The vorticity structure function is calculated using the formula given by \cite{grossmann1992structure} for two-dimensional turbulence as follows
\begin{equation}
    S_\omega = <|\bar{\omega}(\mathbf{x}+\mathbf{r}) - \bar{\omega}(\mathbf{x})|^2>,
\end{equation}
where $<>$ indicates ensemble averaging, $\mathbf{x}$ is the position on the grid, and $\mathbf{r}$ is certain distance from this location. The PDF of the vorticity increment is utilized to assess the capability of SGS models in predicting the coherent vortices in the flow. The vorticity increments at different separations $\mathbf{r}$ is defined as 
\begin{equation}
    \delta \omega (\mathbf{r}) = \omega(\mathbf{x} + \mathbf{r}) - \omega(\mathbf{r}).
\end{equation}

\begin{figure*}[htbp]
\centering
\mbox{\subfigure{\includegraphics[width=0.98\textwidth]{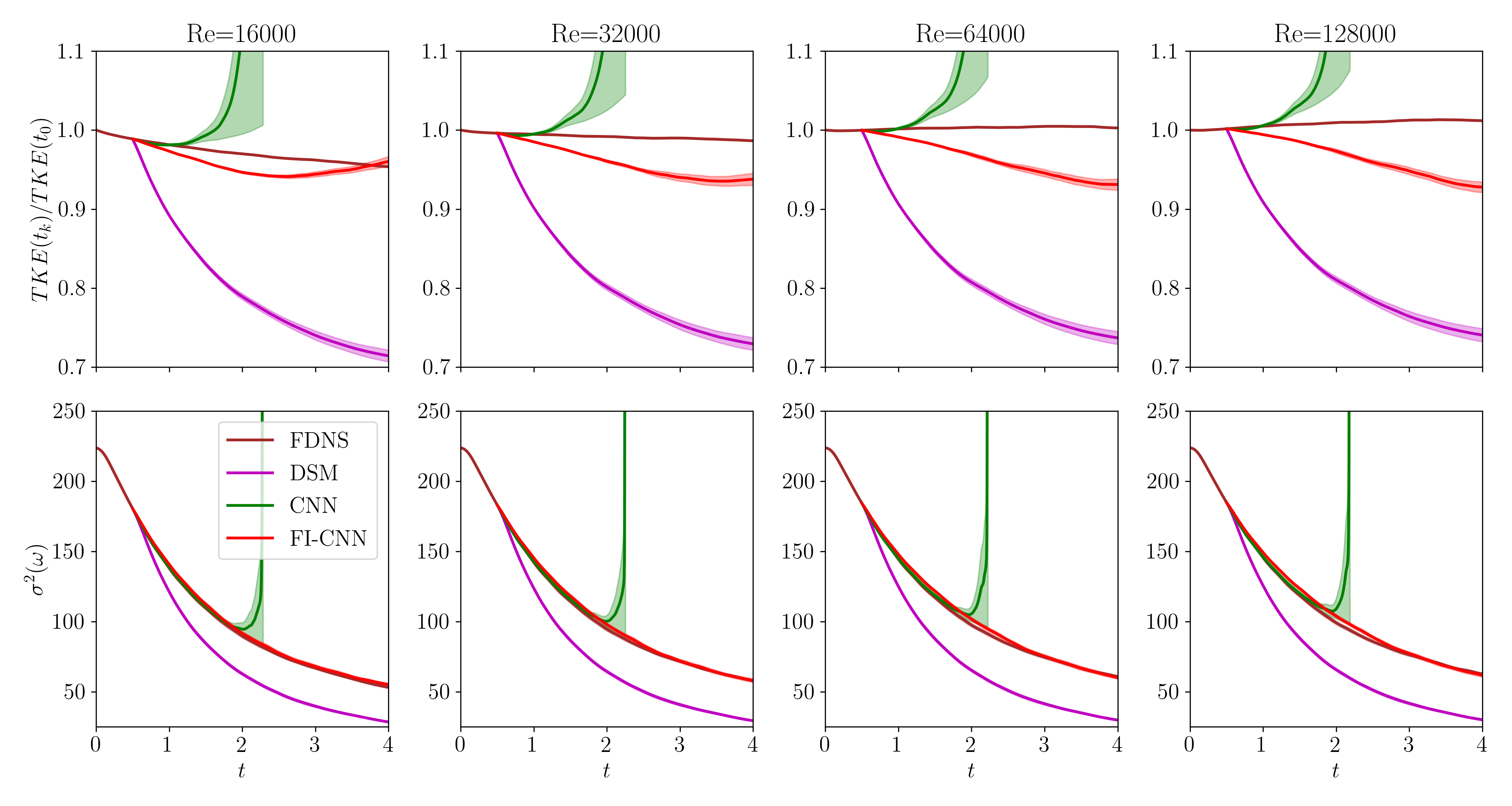}}}
\caption{The time evolution of the turbulent kinetic energy $TKE(t_k)$ normalized by the initial turbulent kinetic energy $TKE(t_0)$ (top row) and vorticity variance (bottom row) for different Reynolds numbers at $256^2$ grid resolution. The solid line shows the mean from LES runs for five different initial conditions and the shaded area corresponds to one standard deviation. The LES simulation starts at $t=0.5$ after the initial spin-up time (i.e., once the turbulence has set in). The CNN and FI-CNN models are trained using the data generated from a single initial condition at Reynolds number Re = 16000. }
\label{fig:sn_turb_stats}
\end{figure*}

We reiterate here that neural network-based SGS models are trained using the data for Reynolds number Re = 16000 and a single initial condition. Once the models are trained, the LES coupled with SGS models is performed for Reynolds number up to Re = 128000 and for five different initial conditions. Fig.~\ref{fig:sn_turb_stats} shows the evolution of turbulent kinetic energy and vorticity variance for LES runs with five different initial conditions and for several Reynolds numbers. For all the LES runs, we initialize the vorticity field at $t=0.5$ after the initial spin-up period using the filtered DNS solution. We can observe that the model $\mathcal{M}_{\text{CNN}}$ is stable only for short time and quickly becomes unstable after $t \approx 2.0$ even for Reynolds number Re = 16000 which was included in the training. In contrast to $\mathcal{M}_{\text{CNN}}$, model $\mathcal{M}_{\text{FI-CNN}}$ is stable for all test cases conducted here without any post-processing of the predicted SGS source term. The ensemble averaging procedure in DSM leads to highly dissipative results and is noticeable in the overprediction of the energy decay rate. The results of the LES runs with $\mathcal{M}_{\text{FI-CNN}}$ have the best agreement with filtered DNS solution for both turbulent kinetic energy and the vorticity variance. 

\begin{figure*}[htbp]
\centering
\mbox{\subfigure{\includegraphics[width=0.98\textwidth]{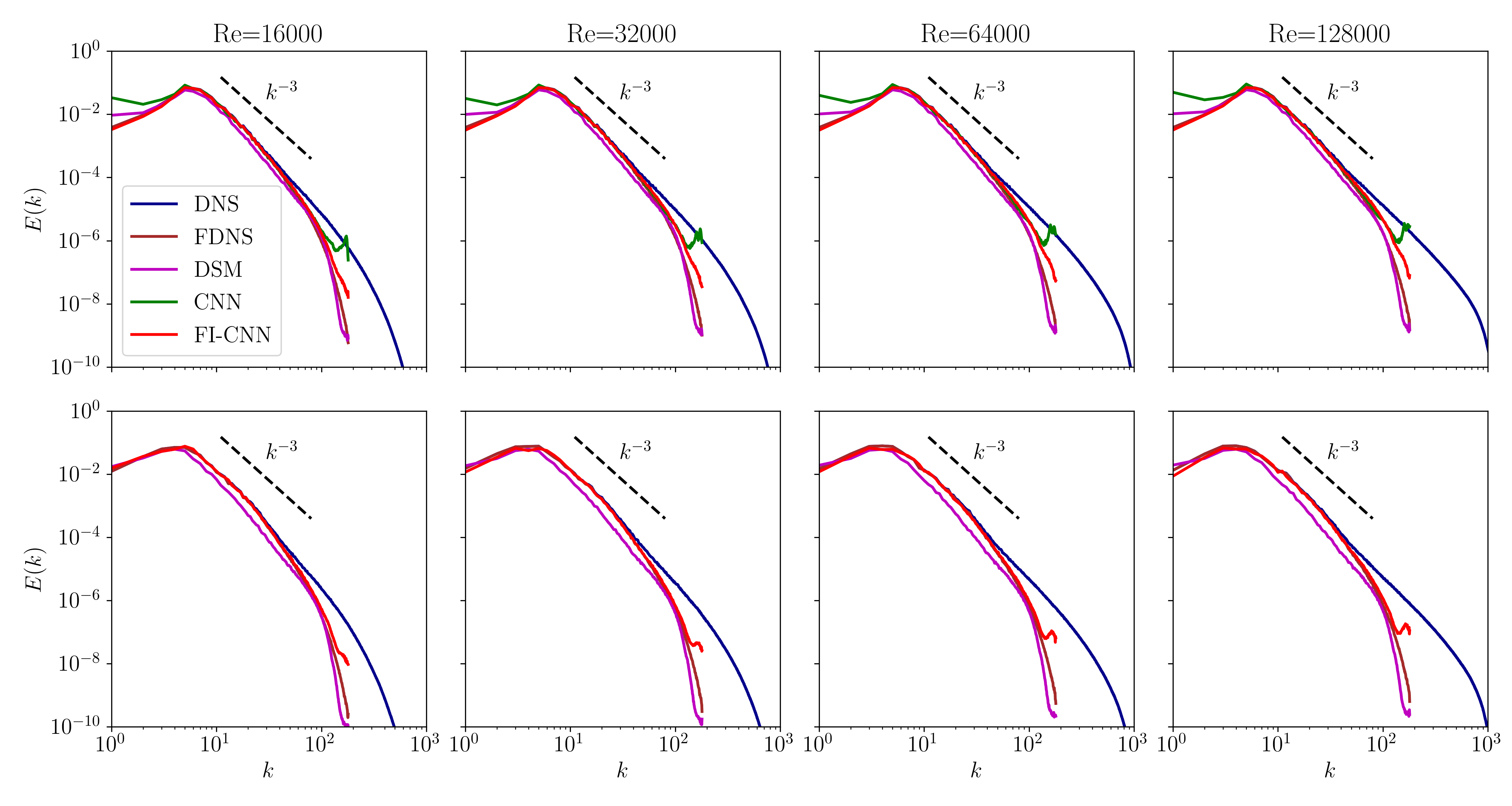}}}
\caption{\textit{A posteriori} kinetic-energy spectra for different Reynolds numbers at $t=2.0$ (top row) and $t=4.0$ (bottom row). These results are obtained from LES runs with five different initial conditions and only mean kinetic energy spectrum is shown. Note here that the CNN model has diverged and the kinetic-energy spectra for the CNN model is missing at the final time $t=4.0$ (bottom row).}
\label{fig:sn_es}
\end{figure*}

\begin{figure*}[htbp]
\centering
\mbox{\subfigure{\includegraphics[width=0.98\textwidth]{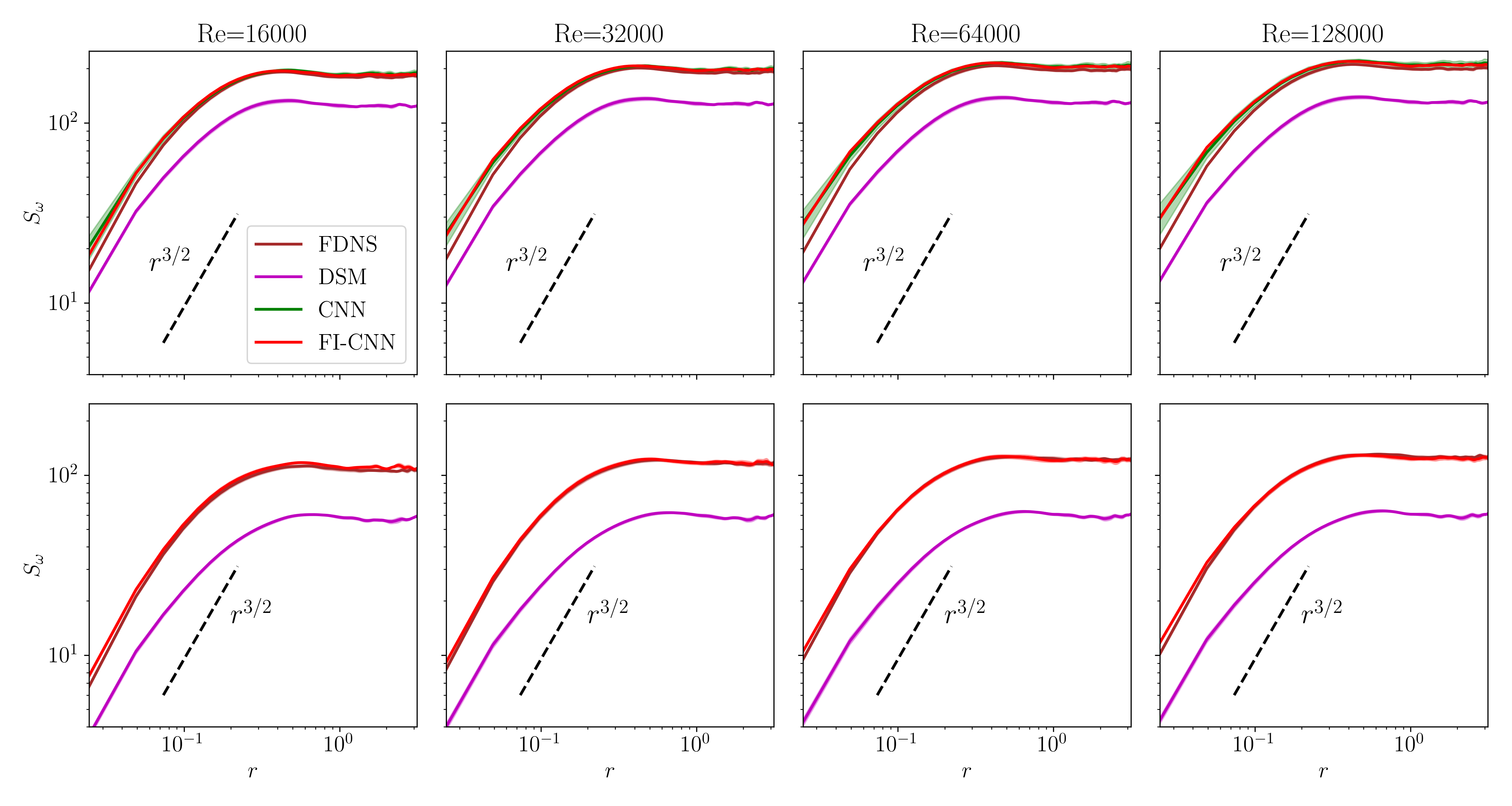}}}
\caption{\textit{A posteriori} second-order vorticity structure for different Reynolds numbers at $t=2.0$ (top row) and $t=4.0$ (bottom row). These results are obtained from LES runs with five different initial conditions and the solid line shows the mean vorticity structure and the shaded area corresponds to one standard deviation. Note here that the CNN model has diverged and the vorticity structure for the CNN model is not present at the final time $t=4.0$ (bottom row).}
\label{fig:sn_vs}
\end{figure*}

Fig.~\ref{fig:sn_es} displays the kinetic-energy spectra at intermediate time $t=2.0$ and at final time $t=4.0$ obtained from LES runs with five different initial conditions for multiple Reynolds number. Although the LES runs coupled with $\mathcal{M}_{\text{CNN}}$ is stable at $t=2.0$, the solution is unphysical as seen by the energy pile up near grid cutoff wavenumbers. This behavior is also illustrated in Fig.~\ref{fig:sn_vs} through a large value of vorticity structure function at $t=2.0$ across all Reynolds numbers. The LES runs with $\mathcal{M}_{\text{CNN}}$ has diverged around $t\approx2.5$ (see Fig.~\ref{fig:sn_turb_stats}), and, therefore the kinetic-energy spectra and vorticity structure function are missing at $t=4.0$ in Fig.~\ref{fig:sn_es} and Fig.~\ref{fig:sn_vs}, respectively. There is a very good agreement between the kinetic-energy spectra for LES runs with $\mathcal{M}_{\text{FI-CNN}}$ and filtered DNS solution, especially in the inertial subrange and $k^{-3}$ theoretical scaling is captured accurately. From Fig.~\ref{fig:sn_vs}, we can see that the model $\mathcal{M}_{\text{FI-CNN}}$ is successful in producing the $r^{3/2}$ scaling \cite{kramer2011structure} for the vorticity structure function at small scales and it gradually flattens near the large scales. The excessive dissipation of DSM is also illustrated in Fig.~\ref{fig:sn_es} and Fig.~\ref{fig:sn_vs} via mismatch between kinetic-energy spectra and vorticity structure function between DSM and filtered DNS solution. The successful performance of LES runs with $\mathcal{M}_{\text{FI-CNN}}$ demonstrates that incorporating frame symmetries as hard constraints has been effective in stabilizing the coarse-grid simulation and in ensuring generalized learning across different initial conditions and  Reynolds numbers.

\begin{figure*}[htbp]
\centering
\mbox{\subfigure{\includegraphics[width=0.98\textwidth]{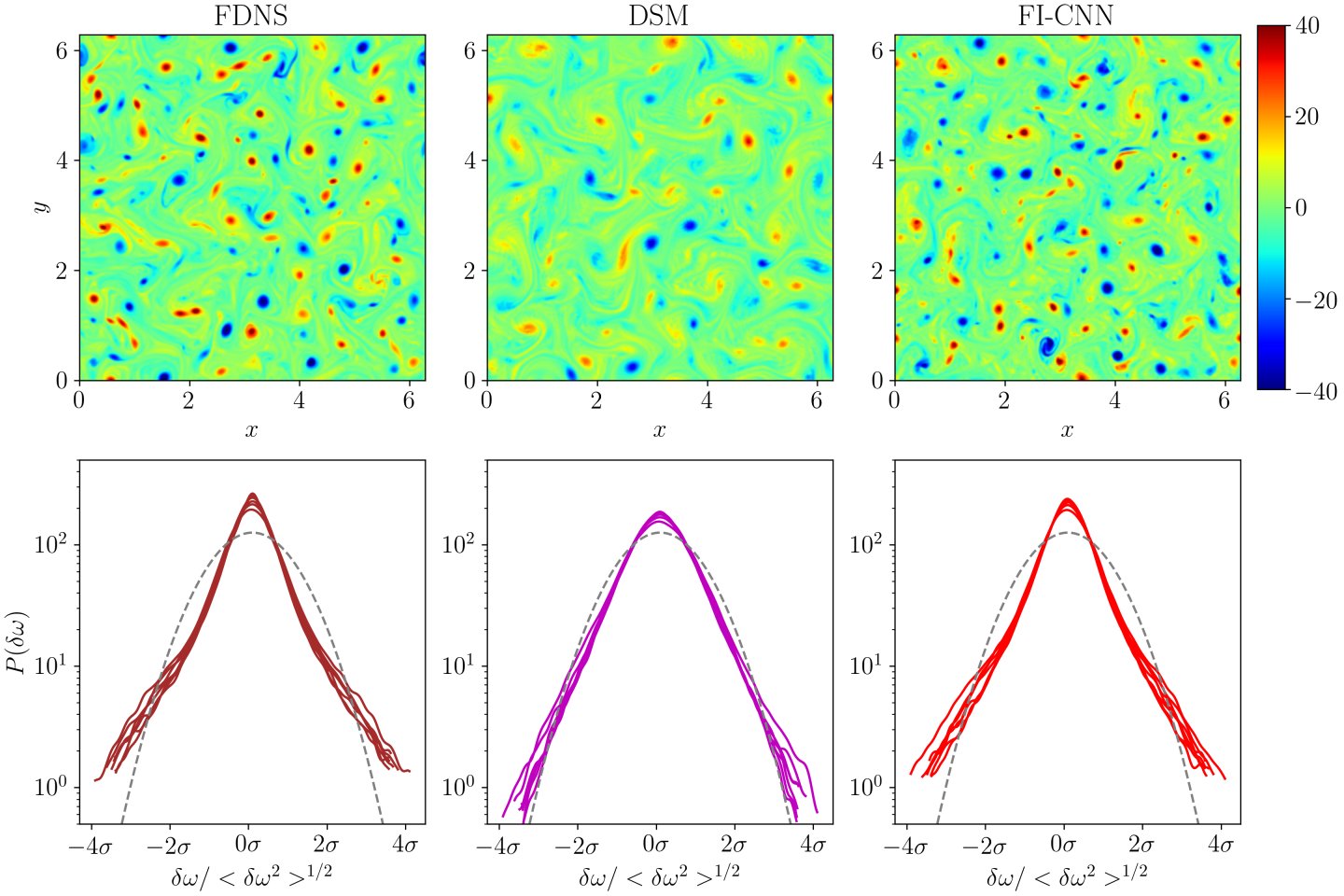}}}
\caption{Snapshots of the vorticity distribution for Re = 16000 taken at final time $t=4.0$ (top row) from different models and compared qualitatively against the FDNS solution. The bottom row displays the probability density function $P(\delta \omega)$ of the vorticity increments $\delta \omega$ for separations $r = 2\pi/256, 2\pi/128, 2\pi/64, 2\pi/32,2\pi/16, 2\pi/8, \text{and} 2\pi/4$ computed from different models at Re = 16000. A Gaussian distribution is given in gray dashed line for comparison. We also note that the plain vanilla CNN becomes numerically unstable and unbounded before $t=4.0$.}
\label{fig:pdf_16}
\end{figure*}

\begin{figure*}[htbp]
\centering
\mbox{\subfigure{\includegraphics[width=0.98\textwidth]{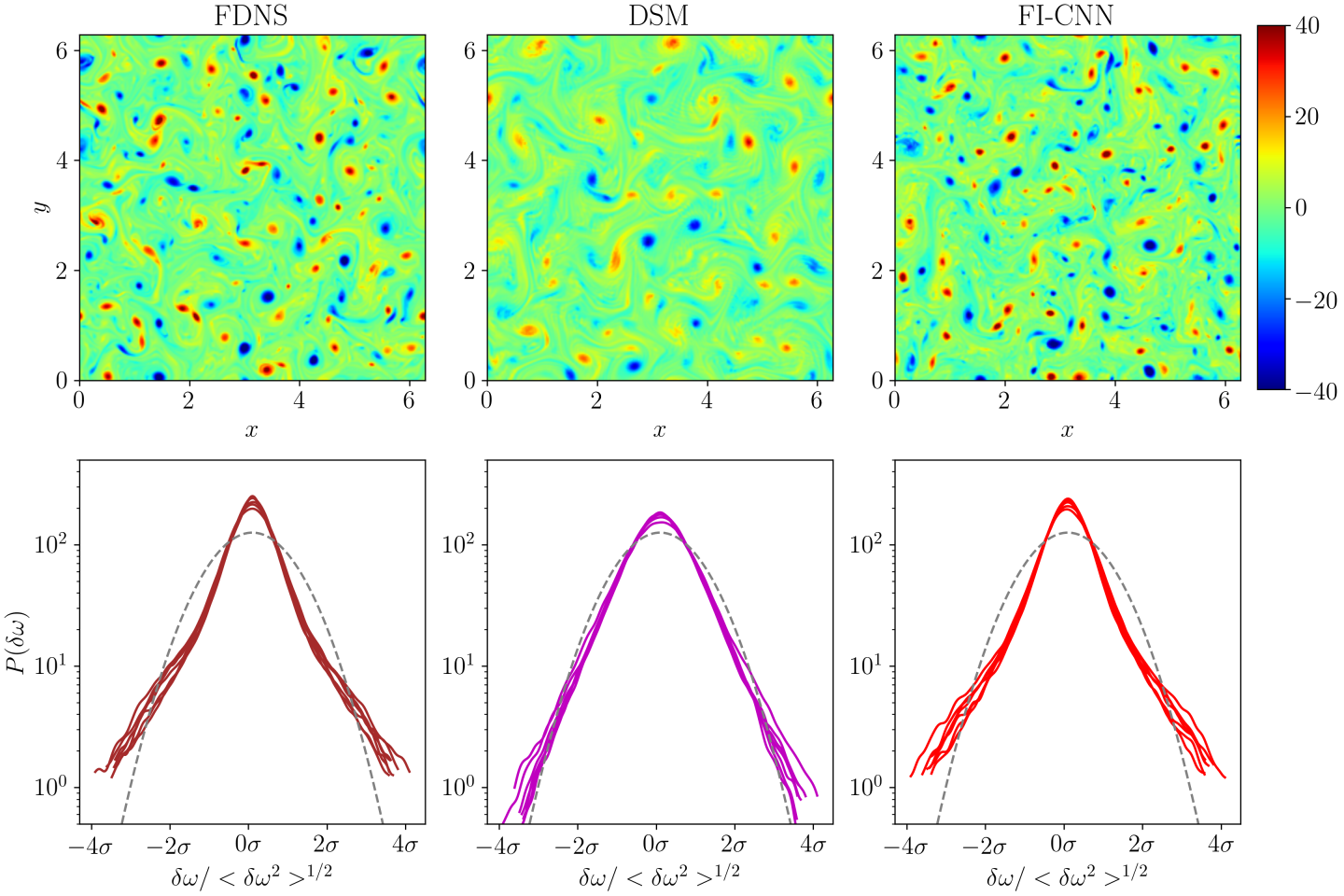}}}
\caption{Snapshots of the vorticity distribution for Re = 32000 taken at final time $t=4.0$ (top row) from different models and compared qualitatively against the FDNS solution. The bottom row displays the probability density function $P(\delta \omega)$ of the vorticity increments $\delta \omega$ for separations $r = 2\pi/256, 2\pi/128, 2\pi/64, 2\pi/32,2\pi/16, 2\pi/8, \text{and} 2\pi/4$ computed from different models at Re = 32000. A Gaussian distribution is given in gray dashed line for comparison. We also note that the plain vanilla CNN becomes numerically unstable and unbounded before $t=4.0$.}
\label{fig:pdf_32}
\end{figure*}

\begin{figure*}[htbp]
\centering
\mbox{\subfigure{\includegraphics[width=0.98\textwidth]{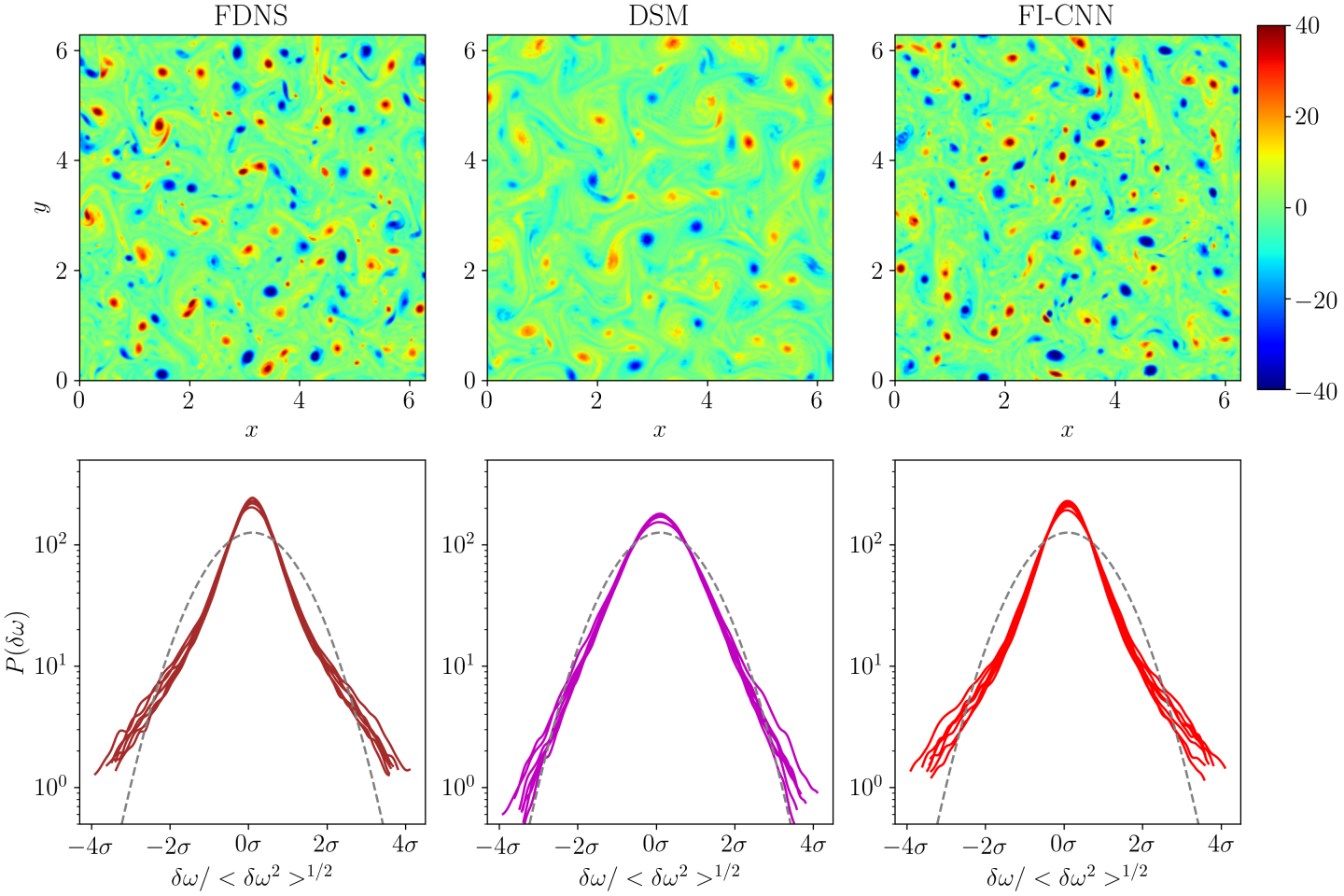}}}
\caption{Snapshots of the vorticity distribution for Re = 64000 taken at final time $t=4.0$ (top row) from different models and compared qualitatively against the FDNS solution. The bottom row displays the probability density function $P(\delta \omega)$ of the vorticity increments $\delta \omega$ for separations $r = 2\pi/256, 2\pi/128, 2\pi/64, 2\pi/32,2\pi/16, 2\pi/8, \text{and} 2\pi/4$ computed from different models at Re = 64000. A Gaussian distribution is given in gray dashed line for comparison. We also note that the plain vanilla CNN becomes numerically unstable and unbounded before $t=4.0$.}
\label{fig:pdf_64}
\end{figure*}

\begin{figure*}[htbp]
\centering
\mbox{\subfigure{\includegraphics[width=0.98\textwidth]{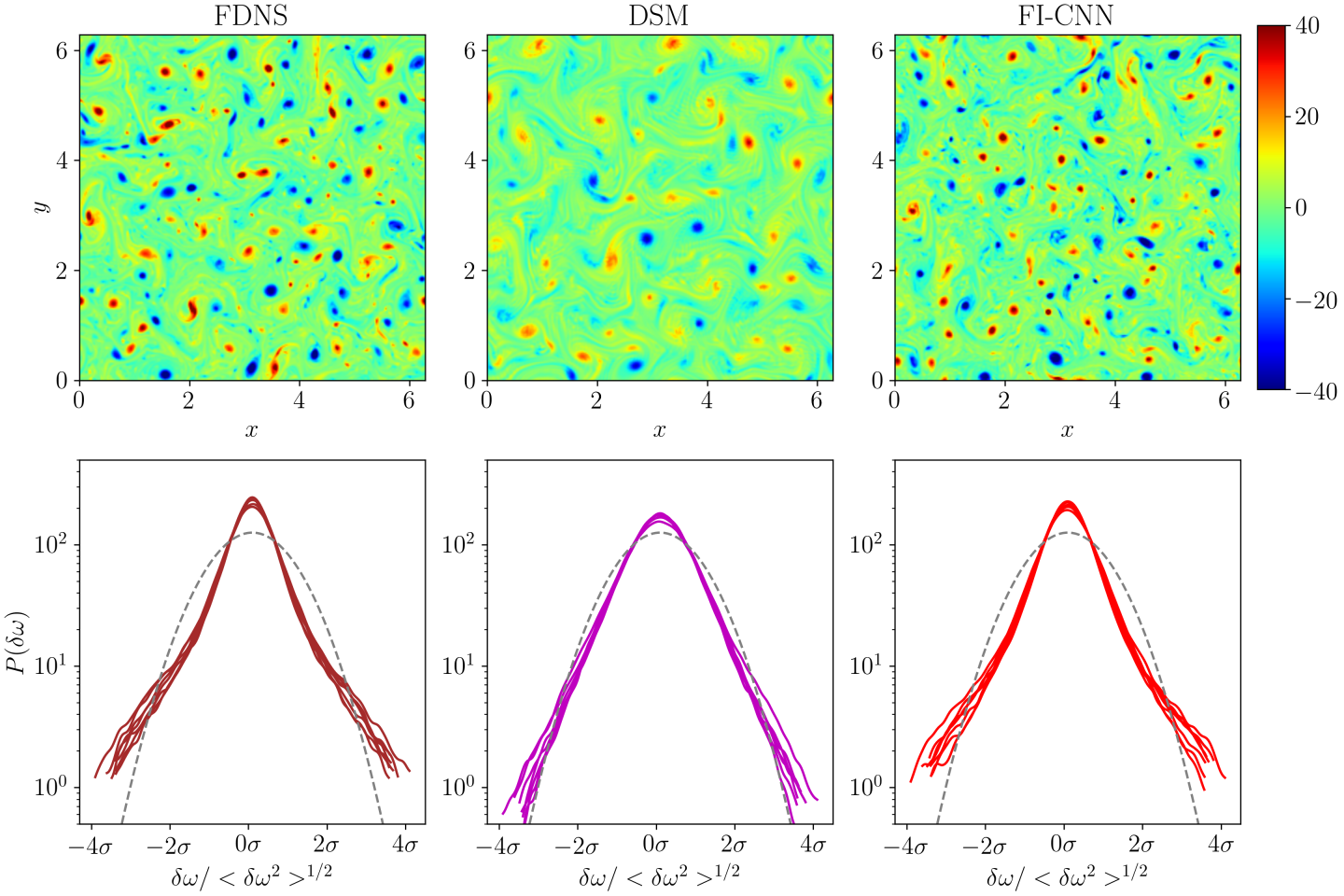}}}
\caption{Snapshots of the vorticity distribution for Re = 128000 taken at final time $t=4.0$ (top row) from different models and compared qualitatively against the FDNS solution. The bottom row displays the probability density function $P(\delta \omega)$ of the vorticity increments $\delta \omega$ for separations $r = 2\pi/256, 2\pi/128, 2\pi/64, 2\pi/32,2\pi/16, 2\pi/8, \text{and} 2\pi/4$ computed from different models at Re = 128000. A Gaussian distribution is given in gray dashed line for comparison. We also note that the plain vanilla CNN becomes numerically unstable and unbounded before $t=4.0$.}
\label{fig:pdf_128}
\end{figure*}

Figs.~\ref{fig:pdf_16}-\ref{fig:pdf_128} provides the visualization of vorticity field and probability density function (PDF) of vorticity increments for Reynolds number Re = 16000 to Re = 128000 computed using the filtered DNS solution, LES with DSM model, and LES with $\mathcal{M}_{\text{FI-CNN}}$ at final time $t=4.0$. We remark here that these results correspond to only one initial condition that is different from the one used for training the neural network-based SGS models. Even though the LES with DSM model is successful in capturing large-scale structures in the flow, it fails to capture the small-scale structure due to excessive dissipation. The LES with $\mathcal{M}_{\text{FI-CNN}}$ is able to capture both large- and small-scale structures in the flow, and this can be ascertained to the stabilizing property of $\mathcal{M}_{\text{FI-CNN}}$ in the \textit{a posteriori} deployment without any post-processing of the predicted SGS source term. Qualitatively, the vorticity field obtained from LES with model $\mathcal{M}_{\text{FI-CNN}}$ is very similar to the filtered DNS solution. The similarity in the shape of the PDF of vorticity increments as shown in Figs.~\ref{fig:pdf_16}-\ref{fig:pdf_128} suggests the scale-invariant statistics of turbulence at all Reynolds numbers investigated in this study. The shape of the PDF of vorticity increments predicted by the LES with $\mathcal{M}_{\text{FI-CNN}}$ matches with the shape of the filtered DNS solution, and the heavy exponential tails in the PDF are related to the presence of coherent vortices in the flow. These heavy exponential tails are missing in the PDF of the solution obtained from LES with DSM, and it follows the Gaussian distribution.

\begin{figure*}[htbp]
\centering
\mbox{\subfigure{\includegraphics[width=0.98\textwidth]{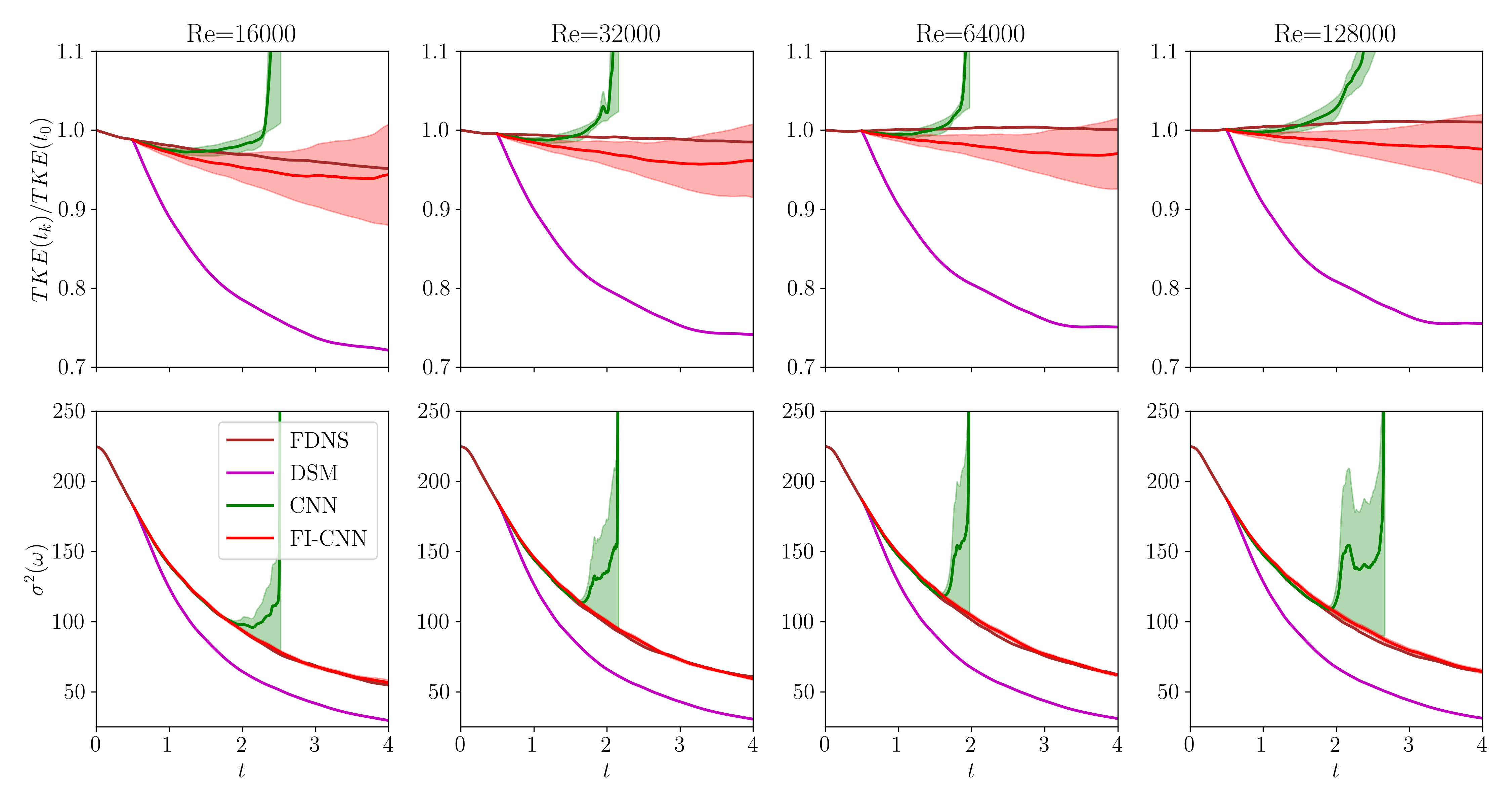}}}
\caption{The time evolution of the turbulent kinetic energy $TKE(t_k)$ normalized by the initial turbulent kinetic energy $TKE(t_0)$ (top row) and vorticity variance (bottom row) for different Reynolds numbers at $256^2$ grid resolution. The LES simulation starts at $t=0.5$ after the initial spin-up time (i.e., once the turbulence has set in). For the CNN and FI-CNN models, an ensemble of neural networks is trained using the data generated from a single initial condition at Reynolds number Re = 16000 with different weights initialization. The solid line shows the mean from LES runs for a single initial condition with different trained networks and the shaded area corresponds to one standard deviation.}
\label{fig:dn_turb_stats}
\end{figure*}

Next, we examine the robustness of the neural network-based SGS model by training an ensemble of neural networks using randomization-based approaches where different random initialization of weights are utilized for generating ensembles. Specifically, we train five neural networks for both models $\mathcal{M}_{\text{CNN}}$ and $\mathcal{M}_{\text{FI-CNN}}$ using the same dataset as discussed in Section~\ref{sec:data}. This method is also applied to quantify the model-form uncertainty in deep learning \cite{lakshminarayanan2017simple}. Fig.~\ref{fig:dn_turb_stats} shows the time evolution of turbulent kinetic energy and vorticity variance at different Reynolds numbers. The time evolution of the TKE in Fig.~\ref{fig:dn_turb_stats} implies that the weights of the neural networks for $\mathcal{M}_{\text{FI-CNN}}$ are learned in such a way that the final models are overall dissipative in nature (as indicated by the solid line for mean from different LES runs). The vorticity variance predicted by the model $\mathcal{M}_{\text{FI-CNN}}$ is more accurate compared to DSM and is very close to the filtered DNS solution.

\begin{figure*}[htbp]
\centering
\mbox{\subfigure{\includegraphics[width=0.98\textwidth]{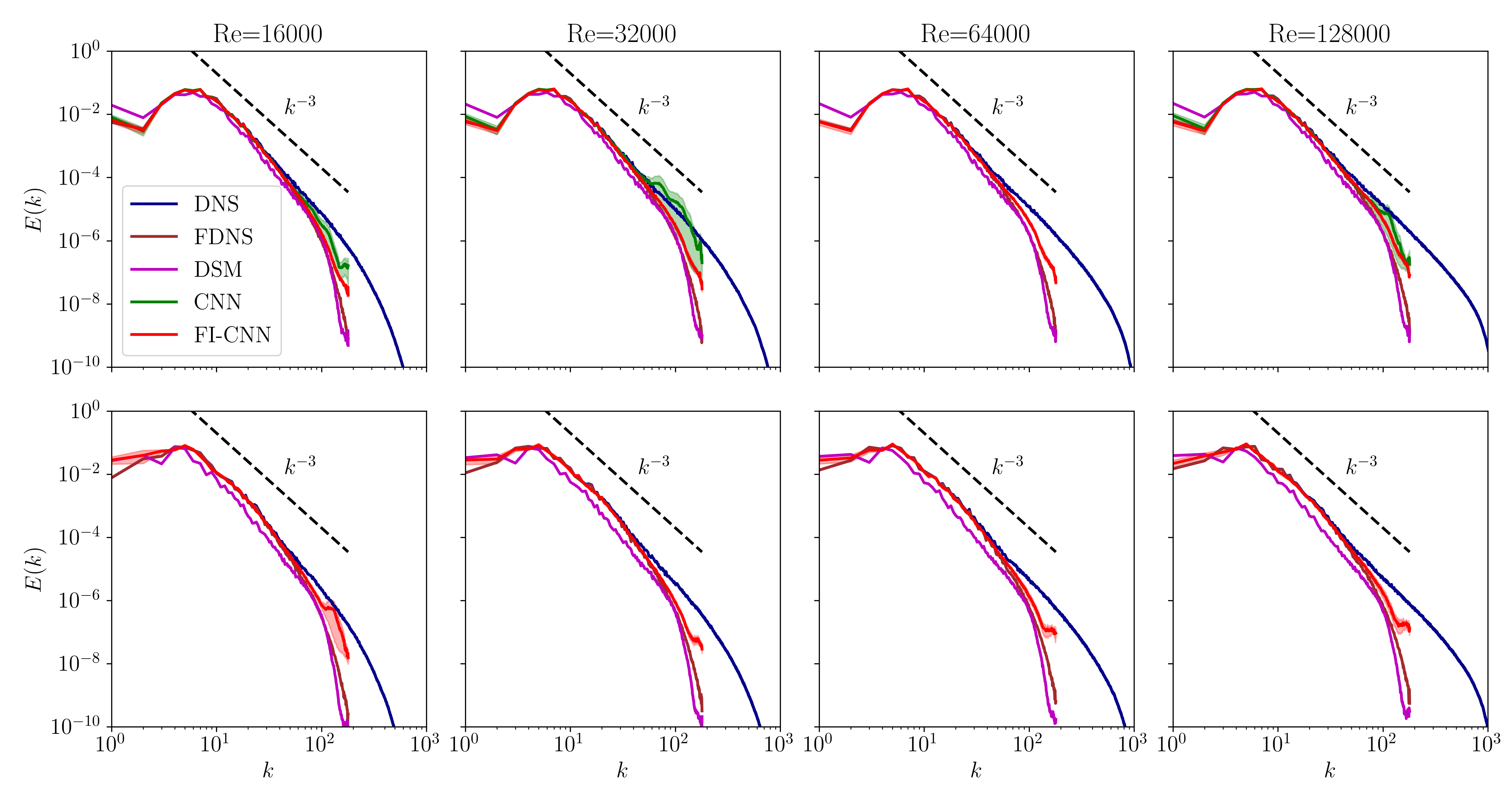}}}
\caption{\textit{A posteriori} kinetic-energy spectra for different Reynolds numbers at $t=2.0$ (top row) and $t=4.0$ (bottom row). The solid line shows the mean from LES runs for a single initial condition with different trained networks and the shaded area corresponds to one standard deviation. Note here that the CNN model has already diverged and the kinetic-energy spectra for the CNN model is missing at the final time $t=4.0$ (bottom row).}
\label{fig:dn_es}
\end{figure*}

\begin{figure*}[htbp]
\centering
\mbox{\subfigure{\includegraphics[width=0.98\textwidth]{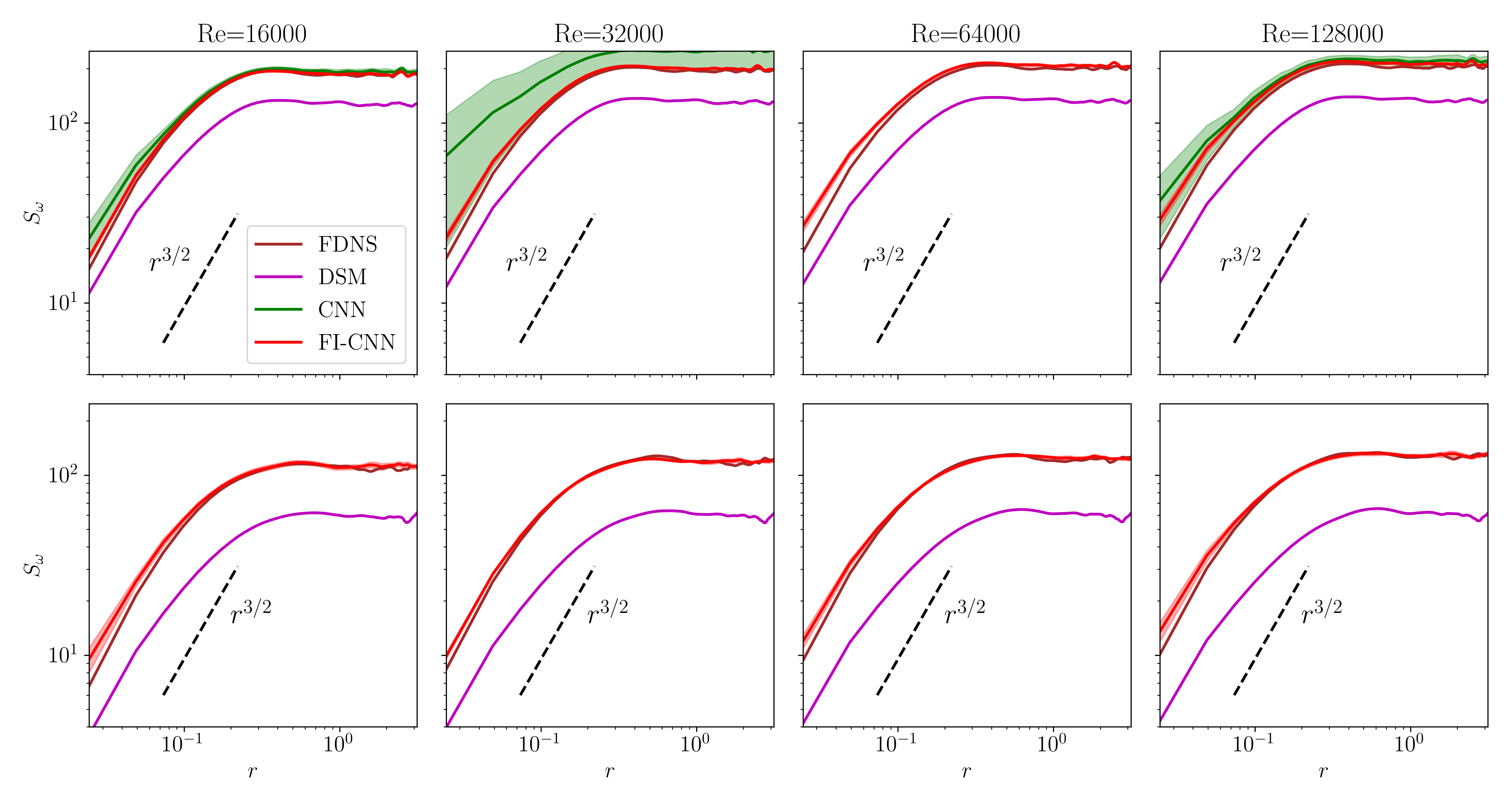}}}
\caption{\textit{A posteriori} second-order vorticity structure for different Reynolds numbers at $t=2.0$ (top row) and $t=4.0$ (bottom row). The solid line shows the mean from LES runs for a single initial condition with different trained networks and the shaded area corresponds to one standard deviation. Note here that the CNN model has diverged and the vorticity structure for the CNN model is not present at the final time $t=4.0$ (bottom row).}
\label{fig:dn_vs}
\end{figure*}

Fig.~\ref{fig:dn_es} depicts the kinetic-energy spectra at intermediate time $t=2.0$ and at final time $t=4.0$ obtained from LES runs for a single initial condition with different network-based SGS models for multiple Reynolds numbers. We observe the energy pile up near grid cutoff wavenumbers for the LES runs coupled with $\mathcal{M}_{\text{CNN}}$ and this suggests that the solution is unphysical. This behavior is also demonstrated in Fig.~\ref{fig:dn_vs} through a large value of vorticity structure function at $t=2.0$ across all Reynolds numbers. The LES runs with $\mathcal{M}_{\text{CNN}}$ has diverged around $t\approx2.5$ (as seen by large $TKE$ in Fig.~\ref{fig:dn_turb_stats}), and, therefore the kinetic-energy spectra and vorticity structure function are missing at $t=4.0$ in Fig.~\ref{fig:dn_es} and Fig.~\ref{fig:dn_vs}, respectively. The kinetic-energy spectra for LES runs with $\mathcal{M}_{\text{FI-CNN}}$ is highly accurate and shows an excellent agreement with the filtered DNS solution, especially in the inertial subrange. The small uncertainty band also suggests that an ensemble of neural networks have produced very similar statistics for $\mathcal{M}_{\text{FI-CNN}}$. Fig.~\ref{fig:dn_vs} shows that the model $\mathcal{M}_{\text{FI-CNN}}$ is successful in capturing the $r^{3/2}$ scaling for the vorticity structure function at small scales and flattening near large scales. With this numerical experiment, we can establish that the $\mathcal{M}_{\text{FI-CNN}}$ is robust, trustworthy, and stable in the LES, and it also ensures generalizable learning across different initial conditions and  Reynolds numbers.

\section{Concluding remarks} \label{sec:conclusion}
Closure modeling in fluid dynamics simulations refers to parameterizing the interactions between high-fidelity and coarse-fidelity descriptions. In this study, we explore data-driven closure modeling strategies to improve both the accuracy and generalizability of such residual models. The motivation behind data-driven closure modeling stems from the fact that most of the existing SGS models are derived based on physical and mathematical considerations, and might not account for the important transfer of kinetic energy from small scales to large scales (i.e., back-scatter) \cite{hewitt2020resolving}. We introduce a frame invariant neural network architecture aiming at embedding physical symmetries directly into the structure of the convolutional neural networks. Thus, our model theoretically guarantees the frame symmetries, including translation, Galilean, and rotation invariance both during training and inference. The embedding of physical symmetries as hard constraints not only improves the accuracy of the model but notably improves the generalization of the model, and eventually makes the model stable in their \textit{a posteriori} deployment without any clipping. 

We test the proposed framework for subgrid-scale modeling of Kraichnan turbulence in \textit{a priori} and \textit{a posteriori} settings. The performance of the proposed framework is evaluated using several metrics like kinetic energy spectra, vorticity structure, and vorticity increments. Based on our analysis, we concluded that symmetry preservation has the potential to improve the accuracy, generalizability, and stability of the SGS model, besides embedding important geometric properties of the underlying PDEs into deep learning models. This work also illustrates a broader lesson on how to combine machine learning with physics for scientific computing. It may be argued that two-dimensional turbulence is far from reality. However, it is generally considered as a canonical testbed for geophysical turbulence in the atmosphere and oceans. Our future development will be focused on scaling up the proposed frame invariant closure modeling framework to solve more realistic three-dimensional turbulent flows. Another interesting avenue is to apply this framework for learning parameterization models for geophysical flows, paving the way for improved weather and climate prediction.     

\section*{Data availability}
The data that supports the findings of this study are available within the article. Implementation details and Python scripts can be accessed from the Github repository \cite{githubml}.

\begin{acknowledgements} This material is based upon work supported by the U.S. Department of Energy, Office of Science, Office of Advanced Scientific Computing Research under Award Number DE-SC0019290. O.S. gratefully acknowledges their support. Direct numerical simulations and neural network training for this project was carried out using resources of the Oklahoma State University High-Performance Computing Center.

Disclaimer: This report was prepared as an account of work sponsored by an agency of the United States Government. Neither the United States Government nor any agency thereof, nor any of their employees, makes any warranty, express or implied, or assumes any legal liability or responsibility for the accuracy, completeness, or usefulness of any information, apparatus, product, or process disclosed, or represents that its use would not infringe privately owned rights. Reference herein to any specific commercial product, process, or service by trade name, trademark, manufacturer, or otherwise does not necessarily constitute or imply its endorsement, recommendation, or favoring by the United States Government or any agency thereof. The views and opinions of authors expressed herein do not necessarily state or reflect those of the United States Government or any agency thereof.
\end{acknowledgements}

\bibliography{references}

\end{document}